\documentclass[a4paper,12pt]{article}
\pdfoutput=1

\usepackage{fullpage}
\usepackage[english]{babel}
\usepackage{amsmath,amssymb,amsfonts,amsthm}
\usepackage{color}
\usepackage{graphicx}
\usepackage[utf8]{inputenc}
\usepackage{dsfont}
\usepackage{subfigure}
\usepackage{collref}

\usepackage{hyperref}
\definecolor{darkred}{rgb}{0.8,0.1,0.1}
\hypersetup{
    colorlinks=true,         % false: boxed links; true: colored links, false is default
    linkcolor=darkred,
    citecolor=blue,
}

\numberwithin{equation}{section}

\DeclareMathOperator{\partialLR}{\stackrel{\leftrightarrow}{\partial}}
\newcommand{\sbr}[1]{{\scriptscriptstyle (#1)}}

\title{Beyond the unitarity bound in AdS\,/\,\texorpdfstring{\boldmath{CFT${}^{}_\text{(A)dS}$}}{CFT((A)dS)}}

 \author{
   Tom\'{a}s Andrade$^{1,}$\thanks{e-mail: \texttt{tandrade@umail.ucsb.edu}} \ \ and \ Christoph F.~Uhlemann$^{2,}$\thanks{e-mail: \texttt{uhlemann@physik.uni-wuerzburg.de}}\\
   \hfill\\ \hfill\\
   ${}^1$\,Department of Physics, UCSB, Santa Barbara, CA 93106, USA\\
   \hfill\\
   ${}^2$\,Institut f\"ur Theoretische Physik und Astrophysik\\
     Universit\"at W\"urzburg,
    Am Hubland, 97074 W\"urzburg, Germany}

\begin{document}
\maketitle
\sloppy

\begin{abstract}
In this work we expand on the holographic description of CFTs on de Sitter (dS) and anti-de Sitter (AdS) spacetimes and
examine how violations of the unitarity bound in the boundary theory are recovered in the bulk physics.
To this end we consider a Klein-Gordon field on AdS$_{d+1}$ conformally compactified such that the boundary is (A)dS$_d$,
and choose masses and boundary conditions such that the corresponding boundary operator violates the CFT unitarity bound.
The setup in which the boundary is AdS$_d$ exhibits a particularly interesting structure, since in this case the boundary itself has a boundary.
The bulk theory turns out to crucially depend on the choice of boundary conditions on the boundary of the AdS$_d$ slices.
Our main result is that violations to the unitarity bound in CFTs on dS$_d$ and AdS$_d$
are reflected in the bulk through the presence of ghost excitations.
In addition, analyzing the setup with AdS$_d$ on the boundary allows us to draw conclusions on multi-layered AdS/CFT-type dualities.
\end{abstract}

\tableofcontents

\section{Introduction}

The AdS/CFT correspondence relates the dynamics of the fields in a $(d{+}1)$-dimensional gravitational theory with asymptotically anti-de Sitter (AdS) boundary conditions
to that of operators in a non-gravitational $d$-dimensional conformal field theory (CFT) on the
boundary \cite{Maldacena:1997re, Witten:1998qj, Gubser:1998bc}.
The prime examples involve gravity on global and Poincar\'e AdS, which are dual to CFTs on the cylinder and the plane, respectively.
There are, however, also cases of interest beyond these dualities with flat-space CFTs, since the study of quantum field theory in curved spacetimes is of general
interest and it is natural to approach it holographically.
Moreover, CFTs on manifolds with boundary (BCFT) \cite{Cardy:1984bb, 1995NuPhB.455..522M} have received attention recently, e.g.\ in the context of brane configurations with
branes ending on branes \cite{Gaiotto:2008sa,Gaiotto:2008ak, Aharony:2011yc}.
In this work we will consider certain aspects of the holographic study of CFTs defined on the maximally symmetric de Sitter (dS) and AdS spacetimes.
This is not only a natural first step from flat to generic curved spacetimes, but also provides a link to
BCFT, since the global AdS on which the CFT is defined is conformally related to half of the Einstein static universe.
Furthermore, the case with AdS on the boundary offers an interesting possibility for multi-layered AdS/CFT dualities.

The holographic description of a CFT on a specific background involves gravity on
an asymptotically-AdS space with that prescribed boundary structure.
The geometries for a dual description of CFTs on dS and AdS have been discussed 
recently in \cite{Marolf2011} and \cite{Aharony:2010ay}, respectively,
and earlier related works can be found in 
\cite{Aharony:2002cx, Balasubramanian:2002am, Cai:2002mr, Karch:2000ct, D'Hoker:2007xy, D'Hoker:2007xz}.
It is sufficient in these cases to choose different coordinates on global AdS$_{d+1}$ such that it is sliced
by (A)dS$_{d}$ hypersurfaces, and perform the conformal compactification adapted to these coordinates.
For the AdS slicing this results in two copies of AdS$_d$ on the boundary and
a single AdS$_{d}$ boundary is obtained by taking a $\mathbb Z_2$ quotient of AdS$_{d+1}$.
The bulk theory then depends on boundary conditions on the hypersurface which is fixed under the $\mathbb Z_2$ action,
and the resulting geometry resembles the general construction for BCFT duals outlined in \cite{Takayanagi:2011zk, Fujita:2011fp}.

Facilitated by the matching of bulk isometries and boundary conformal symmetries,
the AdS/CFT correspondence provides a concrete map between the bulk and boundary Hilbert spaces.
For a free scalar field $\phi$ with mass $m$ on AdS with unit curvature radius
there are in principle two dual operators with conformal dimensions $\Delta_{\pm} = d/2 \pm \sqrt{d^2/4 + m^2}$, up to $1/N$ corrections.
This is related to the fact that solutions to the second-order Klein-Gordon equation are characterized by two asymptotic scalings near the conformal boundary.
Imposing boundary conditions such that the slower/faster fall-off is fixed, which we shall refer to as Dirichlet and Neumann boundary conditions below,  
yields a bulk field dual to an operator of dimension $\Delta_+$/$\Delta_-$ \cite{Klebanov:1999tb, Balasubramanian:1998sn}. Note that the conformal dimensions are 
real so long as the Breitenlohner-Freedman (BF) stability bound $m^2 > -d^2/4=:m_\text{BF}^2$ \cite{Breitenlohner:1982bm, Breitenlohner:1982jf}
is respected.
For $m_\text{BF}^2 < m^2 < m_\text{BF}^2 + 1$ Dirichlet and Neumann boundary conditions yield well-defined 
theories \cite{Breitenlohner:1982bm, Breitenlohner:1982jf},
and in fact even more general boundary conditions can be imposed \cite{Witten:2001ua, Berkooz:2002ug}.
On the other hand, as noted in \cite{Klebanov:1999tb, Balasubramanian:1998sn},
Neumann boundary conditions for $m^2 > m_\text{BF}^2 + 1$ lead to $\Delta_- < d/2-1$,
in conflict with unitarity bounds in the CFT \cite{UBMack,Dobrev:1985qv, Minwalla:1997ka}.
Consequently, the freedom in the choice of boundary conditions was expected to break down for $m^2 > m_\text{BF}^2 + 1$.
This expectation was recently confirmed for global and Poincar\'{e} AdS in \cite{Andrade:2011dg}.
A crucial point is that normalizability of the Neumann modes requires a modification of the
symplectic structure \cite{Compere:2008us},
sacrificing manifest positivity of the associated inner product.
Interestingly, the pathologies in the bulk theory show up in different ways for the two cases.
While in global AdS
the Neumann theories contain ghosts for $m^2 > m_\text{BF}^2 +1$,
such that unitarity in the bulk is explicitly violated,
in Poincar\'e AdS there is no manifest violation of bulk unitarity.
Instead, the 2-point function for the Neumann theories is found to be ill-defined even at large separations.

In this article we take a further step towards a holographic understanding of (A)dS CFTs.
We consider a scalar field with $m^2 \geq m_\text{BF}^2 + 1$ on AdS$_{d+1}$ and choose
coordinates and compactification such that the boundary is (A)dS$_d$.
Imposing Neumann boundary conditions in this mass range is
dual to a CFT on (A)dS$_d$ with an operator of scaling dimension $\Delta \leq d/2-1$.
We will investigate the precise way in which this violation of the CFT unitarity bound is reproduced by the dual bulk theory.
In the setup with the boundary CFT defined on AdS$_d$, the bulk theory depends not only
on the boundary conditions on the AdS$_{d+1}$ conformal boundary, which we refer to
as Neumann$_{d+1}$/Dirichlet$_{d+1}$, but also on the
orbifold boundary conditions and on the boundary conditions on the boundary of the AdS$_d$ slices,
referred to as Neumann$_{d}$/Dirichlet$_{d}$ in the following.
Furthermore, due to the fact that the AdS$_d$ boundary itself has a conformal boundary, the
structure of divergences is more involved than for global or Poincar\'e AdS.
Thus, in order to properly deal with this configuration we have to adapt
the well-established procedure of holographic renormalization \cite{Henningson:1998gx,Balasubramanian:1999re,deHaro:2000xn}.
The choice of Neumann$_d$/Dirichlet$_d$ turns out to be quite crucial.
For Dirichlet$_d$ the adaption of regularization and renormalization is straightforward,
and we find the complete sets of Dirichlet$_{d+1}$ and Neumann$_{d+1}$ modes normalizable with respect to the renormalized inner product.
On the other hand, our construction of the theory with Neumann$_d$ boundary condition leads to a drastically reduced spectrum of normalizable
modes, making the AdS$_{d+1}$ theory equivalent to an AdS$_d$ theory in a trivial way.
This will allow us to draw some conclusions on the possibility of multi-layered holographic dualities,
which were speculated to arise for boundaries with negative cosmological constant in \cite{Compere:2008us}.
The setup with dS on the boundary, on the other hand, is obtained from global AdS by a coordinate transformation
which merely results in a rescaling of the boundary metric, such that this setting is more closely related to global AdS.
However, the dS$_d$ slicing covers only a patch of AdS$_{d+1}$ bounded by a horizon,
analogous to the Lorentzian Poincar\'{e} AdS.
We will investigate whether there is a similarly tricky manifestation of the pathologies as found for Poincar\'{e} AdS in \cite{Andrade:2011dg}.

The paper is organized as follows.
In section~\ref{sec:slicings} we introduce the setups for a holographic description
of CFTs on (A)dS and give the relevant properties of the Klein-Gordon field in these settings.
Unitarity of the bulk theories for AdS$_d$ and dS$_d$ on the boundary is studied in sections \ref{sec:AdS-on-boundary}
and \ref{sec:dS-on-boundary}, respectively,
and we conclude in section~\ref{sec:conclusions}.
In an appendix we discuss a scalar field with tachyonic mass below the BF bound on global AdS.

\section{(A)\texorpdfstring{dS$_\text{d}$}{dS\_d} slicings of A\texorpdfstring{dS$_\text{d+1}$}{dS\_d+1}}\label{sec:slicings}
\begin{figure}[htb]
\center
\subfigure[][]{
\includegraphics[width=0.33\linewidth]{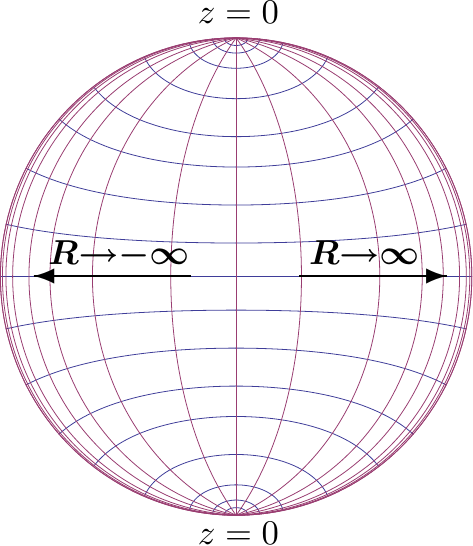}\label{fig:AdSslicing}
}\qquad\qquad
\subfigure[][]{
\includegraphics[width=0.28\linewidth]{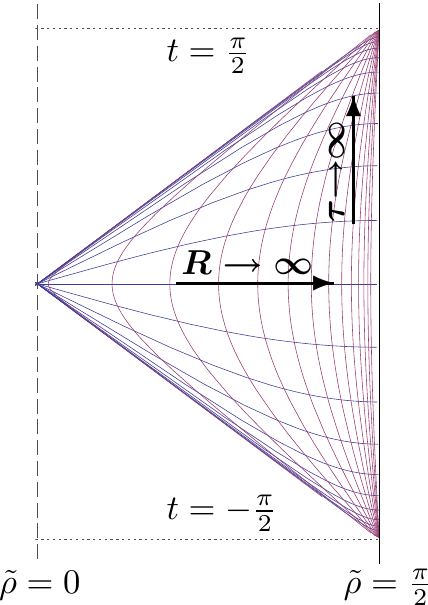}\label{fig:dSslicing}
}
\caption{\label{fig:Slicings}The slicing of global AdS$_{d+1}$ by (A)dS$_d$ hypersurfaces.
         \ref{fig:AdSslicing} shows the Poincar\'e disk representation of  AdS$_{d+1}$  sliced by AdS$_{d}$
         in the $(R,z)$ coordinates used in (\ref{eqn:metric-gen}), (\ref{eqn:ads-d}).
         Horizontal/vertical curves have constant $z$/$R$.
         The boundary consists of two copies of AdS$_{d}$ joined at their boundaries at $z\to 0$.
         \ref{fig:dSslicing} shows the dS$_d$ slicing (\ref{eqn:metric-gen}), (\ref{eqn:ads-ds}) of AdS$_{d+1}$ as cylinder
         with radial coordinate $\tilde\rho=\arctan\rho$ and the $\Omega_{d-1}$ part suppressed.
         Horizontal/vertical curves have constant $\tau$/$R$.
        }
\end{figure}

In this section we introduce the foliations of AdS that will be relevant for the subsequent analysis
and discuss some generic features of the Klein-Gordon field in these coordinates.
We consider AdS$_{d+1}$ with curvature radius $L$
in global coordinates ${(\rho,\zeta,t)\in [0,\infty) {\times} [0,\pi] {\times} \mathbb R}$ such that the line element takes the form
\begin{equation}\label{eqn:ads-global-metric}
 ds^2=-\big(1+\rho^2/L^2\big)dt^2+\frac{1}{1+\rho^2/L^2}d\rho^2+\rho^2 d\Omega_{d-1}^2~,\qquad d\Omega_{d-1}^2=d\zeta^2+\sin^2\zeta d\Omega_{d-2}^2~.
\end{equation}
In the following we discuss coordinate transformations resulting in a metric of the form
\begin{equation}\label{eqn:metric-gen}
  ds^2=dR^2+\lambda(R)^2\,\gamma_{\mu\nu}dx^\mu dx^\nu~,
\end{equation}
with $R\in[0,\infty)$ and the conformal boundary of AdS  at $R=\infty$.
The slicing by dS$_d$ hypersurfaces with Hubble constant $H$ is obtained by the coordinate transformation
$(\rho,t,\Omega_{d-1})\rightarrow(R,\tau,\Omega_{d-1})$ with $\tau\in\mathbb R$ and
\begin{equation}\label{eqn:coord-transf-dS}
 \rho=L \cosh (H\tau)\sinh\frac{R}{L}~, \qquad \tan(t)=L \sinh(H \tau)\tanh\frac{R}{L}~.
\end{equation}
The resulting metric is of the form (\ref{eqn:metric-gen}) with
\begin{equation}\label{eqn:ads-ds}
 \gamma^\text{dS}_{\mu\nu} dx^\mu dx^\nu=-d\tau^2 + H^{-2}\cosh^2(H\tau)\, d\Omega_\text{d-1}^2~,
 \qquad
 \lambda^{}_\text{dS}(R)=L H\sinh\frac{R}{L}~.
\end{equation}
Note that (\ref{eqn:coord-transf-dS}) implies $|\tan(t)/\rho|< 1$, which restricts the
range of $t$ to ${|t|< \arctan(\rho)<\pi/2}$.
The coordinates $(R,\tau)$ therefore cover a patch as shown in figure~\ref{fig:dSslicing}.
The patch is bounded by a causal horizon at $|\tan(t)/\rho| \rightarrow 1$,
which is an infinite-redshift surface as $\lambda_\mathrm{dS}^2$ vanishes there.
The conformal boundary of the patch at $R\rightarrow\infty$ is part of the AdS$_{d+1}$ conformal boundary,
and from (\ref{eqn:ads-ds}) we see that the boundary metric at $R=\infty$ is that of global dS$_d$,
as desired.

The foliation of AdS$_{d+1}$ by AdS$_d$ hypersurfaces with curvature radius $l$
is obtained from the transformation
$(\rho,\zeta,t,\Omega_{d-2})\rightarrow(R,z,\tau,\Omega_{d-2})$
with $z\in(0,\pi/2]$, $\tau\in \mathbb R$ and
\begin{equation}\label{eqn:coord-transf-ads}
 \frac{\rho^2}{L^2}=\csc^2z\cosh^2 \frac{R}{L}-1~,\qquad
 \rho^2\sin^2\zeta=L^2\cot^2z \cosh^2 \frac{R}{L}~,\qquad
 t=L\tau~.
\end{equation}
The resulting metric again is of the form (\ref{eqn:metric-gen}) but with
\begin{equation}\label{eqn:ads-d}
  \gamma^\text{AdS}_{\mu\nu}dx^\mu dx^\nu=\frac{l^2}{\sin^2 z}\big(-d\tau^2+dz^2+\cos^2z\, d\Omega_{d-2}^2\big)~,
  \qquad
  \lambda^{}_\text{AdS}(R)=\frac{L}{l}\cosh\frac{R}{L}~.
\end{equation}
As we have to choose the domain for the sine in the 2$^\text{nd}$ equation in (\ref{eqn:coord-transf-ads}) to be either
$\zeta\in[0,\pi/2)$ or $\zeta\in(\pi/2,\pi]$
we need two patches to cover the full AdS$_{d+1}$.
The patches are `joined' at $\zeta=\pi/2$, the equator of $S^{d-1}$.
This is realized in \cite{Aharony:2010ay} by letting $R$ run on $(-\infty,\infty)$
and choosing the appropriate domains for $\zeta$ on the two half lines.
To obtain a holographic description of a CFT on a single copy of AdS$_d$ we consider
the $\mathbb Z_2$ quotient of global AdS identifying the two patches,
as discussed in \cite{Aharony:2010ay}.
This quotient is covered by the coordinates discussed above for any of the two choices for the domain of $\zeta$.
In turn, this implies that the fields under consideration should have definite $\mathbb Z_2$ parity,
which imposes boundary conditions at $R=0$, as will be discussed in section~\ref{sec:BCandRenormalization}.
Furthermore, note that the resulting single copy of AdS$_d$ at the conformal boundary of AdS$_{d+1}$ has itself a conformal boundary,
which, in the coordinate system (\ref{eqn:ads-d}), corresponds to the locus $z=0$.

The setup for a holographic description of CFTs on AdS$_d$ discussed above resembles
the holographic description for generic BCFT proposed in \cite{Takayanagi:2011zk, Fujita:2011fp}.
For a CFT on a $d$-dimensional manifold $M$ with boundary it was
proposed there to consider as dual a gravitational theory on a $d{+}1$-dimensional asymptotically-AdS
manifold with conformal boundary $M$ and an additional boundary $Q$, such that $\partial Q=\partial M$.
The AdS$_d$ slice at $R\rightarrow\infty$ in our setup corresponds to $M$,
the $\mathbb Z_2$-fixed hypersurface at $R=0$ to $Q$, and
imposing even/odd $\mathbb Z_2$ parity translates to Neumann/Dirichlet boundary conditions on $Q$.
This similarity of the setups can be understood as a consequence of
the relation of CFTs on AdS to BCFTs discussed in the introduction.

\subsection{Klein-Gordon field}\label{sec:KGfield}
We consider a free, massive Klein-Gordon field on AdS$_{d+1}$ foliated by (A)dS$_d$
and discuss the features that apply to both slicings in parallel.
Our starting point is the `bare' bulk action for a free scalar field,
\begin{equation}\label{eqn:KGaction}
 S=-\frac{1}{2}\int d^{d+1}x\, \sqrt{g} \Big(g^{MN}\partial_M\phi \partial_N\phi+m^2\phi^2\Big)~,
\end{equation}
which will later be augmented by
boundary terms.
For a metric of the form (\ref{eqn:metric-gen}) the resulting Klein-Gordon equation reads
\begin{equation}\label{eqn:KG}
\partial_R^2\phi+d\, \frac{\lambda^\prime(R)}{\lambda(R)}\,\partial_R\phi+\lambda(R)^{-2}\:\square_\gamma\phi=m^2\phi~.
\end{equation}
We separate the radial and transverse parts by choosing the ansatz
$\phi(x,R)=\varphi(x)f(R)$, such that $\varphi$ are the modes
on the (A)dS$_d$ slices and $f$ are the radial modes.
Introducing $M$ as separation constant,
(\ref{eqn:KG}) separates into the radial equation
\begin{equation}\label{eqn:radial-eq}
 f^{\prime\prime}+d \,\frac{\lambda^\prime}{\lambda} \,f^\prime=\big(m^2-M^2\lambda^{-2}\big)f~,
\end{equation}
and the (A)dS$_d$ hypersurface part $\square_\gamma\varphi=M^2\varphi$.
The latter is a Klein-Gordon equation for the transverse part with `boundary mass' $M$.
Note that (\ref{eqn:radial-eq}) can be written in Sturm-Liouville form,
\begin{equation}\label{SL PP}
    L f = \alpha f \ , \ \ \ \ \ \ \ {\rm where} \ \      L = \frac{1}{w(R)}\left[ -\frac{d}{dR} \left(p(R) \frac{d}{dR} \right)  + q(R)  \right]~.
\end{equation}
Fixing $p(R) = \lambda(R)^d$, $w(R) = \lambda(R)^{d-2}$ and $q(R) = m^2 \lambda(R)^d$ reproduces (\ref{eqn:radial-eq})
with ${\alpha=M^2}$.
The inner product defined from the `bare' symplectic current associated to (\ref{eqn:KGaction}) is the standard Klein-Gordon product
\begin{equation} \label{eqn:KG-inner-product}
\langle\delta_1\phi,\delta_2\phi\rangle^{}_{\mathcal M}=-i\int_\Sigma d^dx^{}_\Sigma\,\sqrt{g^{}_\Sigma} n^M\,\delta_1\phi^\ast\partialLR^{}_M\delta_2\phi~.
\end{equation}
Choosing for $\Sigma$ an (A)dS$_d$ slice, such that ${n^M=(0,n^\mu)}$, the inner product (\ref{eqn:KG-inner-product}) can also be factorized.
In fact, with $n^\mu=:\lambda(R)^{-1} n_\gamma^\mu$ such that $n_\gamma^\mu$ is normalized w.r.t.\ $\gamma_{\mu\nu}$,
(\ref{eqn:KG-inner-product}) becomes
\begin{equation}\label{eqn:inner-prod-factorized}
 \langle\phi_1,\phi_2\rangle^{}_{\mathcal M} = \langle\varphi_1,\varphi_2\rangle^{}_\mathrm{slice}\, \langle f_1,f_2\rangle^{}_\mathrm{SL}~,
\end{equation}
where $\langle\varphi_1,\varphi_2\rangle^{}_\mathrm{slice}$ is the Klein-Gordon inner product on the (A)dS$_d$ slice
and $\langle f_1,f_2\rangle^{}_\mathrm{SL}$ is the Sturm-Liouville inner product
\begin{align}\label{eqn:inner-prod-factorized2}
 \langle\varphi_1,\varphi_2\rangle^{}_\mathrm{slice}&=
   -i\int_{\partial\Sigma}\! d^{d-1}x^{}_{\partial\Sigma}\, \sqrt{\gamma^{}_{\partial\Sigma}} n_\gamma^\mu\big(\varphi_1^\ast\partialLR^{}_\mu\varphi_2\big)~,
& \langle f_1,f_2\rangle^{}_\mathrm{SL}&=\int_{0}^\infty \!\!dR\; \lambda^{d-2} f_1^\ast f_2~.
\end{align}
Using partial integration and (\ref{eqn:radial-eq}) yields\footnote{
Although the derivation of (\ref{eqn:reducedSL}) is only valid for $M_1^\ast\neq M_2$,
(\ref{eqn:reducedSL}) can be continued to $M^\ast_1 = M_2$ by taking the appropriate limits, as we discuss later.
We also note that for continuous boundary mass (\ref{eqn:reducedSL}) has to be understood in the distributional sense.
This procedure is justified by the fact that the obtained results exhibit conservation and finiteness of the symplectic structure.
}
\begin{equation}\label{eqn:reducedSL}
 \langle f_1,f_2\rangle^{}_\mathrm{SL}=\frac{1}{M^{\ast 2}_1-M_2^2}\lim_{a\rightarrow 0, b\rightarrow\infty}\left[\lambda^d\,\big(f_1^\ast f_2^\prime-{f_1^\prime}^\ast f_2\big)\right]_a^b~.
\end{equation}
The inner product (\ref{eqn:KG-inner-product}) is finite and conserved
for Dirichlet and Neumann boundary conditions if $m^2 < m_\text{BF}^2 + 1$.
However, for larger masses, the holographic renormalization of the bulk action
introduces derivative terms on the boundary, which in turn induce a renormalization of
the inner product \cite{Compere:2008us}.
We shall discuss this issue in detail in section~\ref{sec:AdS-on-boundary}.

\subsection{Asymptotic solutions} \label{sec:asympt}
The covariant boundary terms introduced by the holographic renormalization of the
bulk theory are crucial for the construction of the renormalized inner product. The construction of these terms
involves the asymptotic expansion of the on-shell bulk field, which we shall now discuss.
The relevant computations are most conveniently carried out with the metric in Fefferman-Graham form.
For the dS$_d$ slicing
(\ref{eqn:metric-gen}), (\ref{eqn:ads-ds})
this form is obtained by the coordinate transformation
$y:=2 H^{-1}e^{-R/L}\in(0,2 H^{-1}]$, resulting in the metric
\begin{align}
 ds^2=\frac{L^2}{y^2}\Big(dy^2+\Big(1-\frac{H^2y^2}{4}\Big)^2\,\gamma^\mathrm{dS}_{\mu\nu} dx^\mu dx^\nu\Big)~.
\end{align}
Likewise, for the AdS$_d$ slicing (\ref{eqn:metric-gen}), (\ref{eqn:ads-d}) the transformation $y:=2 l e^{-R/L}\in(0,2l]$ yields
\begin{align}
\label{eqn:ads-ads-fg}
 ds^2=\frac{L^2}{y^2}\Big(dy^2+\Big(1+\frac{y^2}{4l^2}\Big)^2\,\gamma^\mathrm{AdS}_{\mu\nu}dx^\mu dx^\nu\Big)~.
\end{align}
The conformal boundary of AdS$_{d+1}$ is at $y=0$ in both cases.
The asymptotic expansion of $\phi$ in these coordinates is obtained by solving the
Klein-Gorden equation expanded around the conformal boundary.
With $m^2 L^2=:-\frac{d^2}{4}+\nu^2$ we obtain
\begin{equation}\label{eq:asymptSol}
 \phi(x^\mu,y) = y^{\frac{d}{2}-\nu}\phi^{}_\mathrm{D}(x^\mu,y)+y^{\frac{d}{2}+\nu}\phi^{}_\mathrm{N}(x^\mu,y)~,
\end{equation}
where $\phi^{}_\mathrm{N/D}$ have regular power-series expansions around $y=0$,
and in particular
\begin{subequations}\label{eqn:asymptField}
\begin{align}
 \phi^{}_\mathrm{D}&=\phi^\sbr{0}_{\text{D}}+y^2\phi^\sbr{2}_\mathrm{D}+\dots~,&
\phi^\sbr{2}_\mathrm{D}&=\frac{1}{4(\nu-1)}\square^W_{\gamma}\phi^\sbr{0}_\mathrm{D}~,& \nu&\in(1,2)~,\\
\phi^{}_\mathrm{D}&=\phi^\sbr{0}_\mathrm{D}+y^2\log(y) \phi^{\sbr{2}}_\mathrm{D}+\dots~,&
 \phi^\sbr{2}_\mathrm{D}&=-\frac{1}{2}\square^W_{\gamma}\phi^\sbr{0}_\mathrm{D}~,
 & \nu &=1~.
\end{align}
\end{subequations}
Here we have defined $\square^W_{\gamma}:=\square_{\gamma}-\frac{d-2\nu}{4(d-1)} R[\gamma]$~,
with the curvature of the hypersurface metric $R[\gamma]$.
The curvature convention is such that
$\mathcal R[\gamma^\text{AdS}]=-l^{-2}d(d-1)$ and $\mathcal R[\gamma^\text{dS}]=H^2 d(d-1)$.

\section{AdS on the boundary}  \label{sec:AdS-on-boundary}

In this section we study the case of AdS on the boundary. After setting up the regularization
and renormalization procedure we discuss the
Dirichlet$_d$ theory in the mass range dual to a CFT beyond the unitarity bound and
discuss the special properties of the Neumann$_d$ theories.

\subsection{Renormalization and boundary conditions} \label{sec:BCandRenormalization}
\begin{figure}[htbp]
\begin{center}
\subfigure[][]{ \label{fig:adsads-bndy}
  \includegraphics[width=0.26\linewidth]{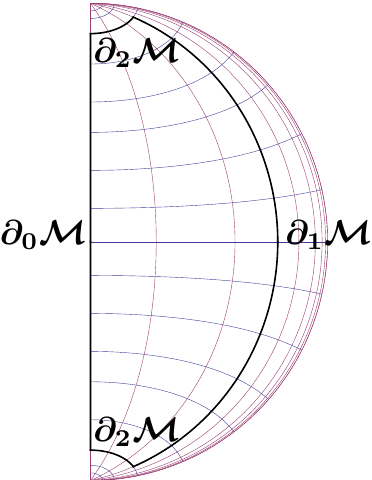}
}\qquad\qquad
\subfigure[][]{ \label{fig:dsads-bndy}
  \includegraphics[width=0.22\linewidth]{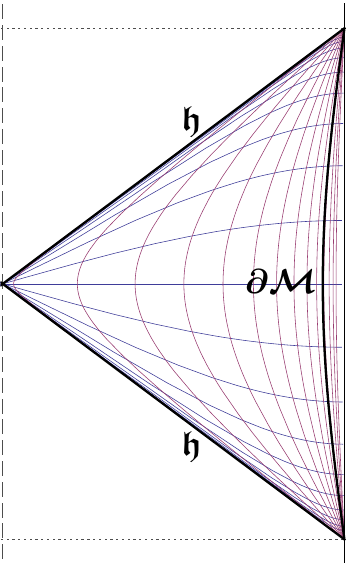}
}
\caption{The boundaries of the regularized geometries.
         \ref{fig:adsads-bndy} shows the regularization of the AdS slicing
         discussed in section~\ref{sec:BCandRenormalization}.
         The intersection of $\partial_1\mathcal M$ and $\partial_2\mathcal M$ is $\partial\partial\mathcal M$.
         \ref{fig:dsads-bndy} shows the regularization of the dS slicing
         discussed in section~\ref{sec:dS-on-boundary}.
         \label{fig:regularizedSlicings}}
\end{center}
\end{figure}

We consider the AdS$_d$ slicing of AdS$_{d+1}$
using the coordinates $(y,z,\tau,\Omega)$ such that the metric is of the Fefferman-Graham form
(\ref{eqn:ads-ads-fg}).
The action (\ref{eqn:KGaction}) evaluated on-shell is divergent as a power series in a vicinity of the boundary at $y=0$,
and we also expect divergences from $z=0$.
To renormalize the divergences we introduce cut-offs at $y=\epsilon_1$, $z=\epsilon_2$
and boundary counterterms to render the asymptotic expansions in $y$, $z$ finite
as the cut-offs are removed by $\epsilon_{1/2}\rightarrow 0$.
The form of the cut-offs is the standard prescription adapted to the current slicing,
and is illustrated in figure~\ref{fig:adsads-bndy}.
We use the notation $\mathcal M=\text{AdS}_{d+1}$
and parametrize the boundary $\partial \mathcal M$ of the regularized $\mathcal M$ as follows:
$\partial^{}_0\mathcal M:=\lbrace\mathcal M \,\vert\, y=2l, z>\epsilon_2\rbrace$
is the hypersurface which is invariant under the orbifold action,
$\partial^{}_1\mathcal M:=\lbrace\mathcal M \,\vert\, y=\epsilon_1, z>\epsilon_2\rbrace$
denotes the (regularized) AdS$_d$ part at large $R$,
$\partial^{}_2\mathcal M:=\lbrace\mathcal M \,\vert\, \epsilon_1<y<2l, z=\epsilon_2\rbrace$
consists of the boundaries of the AdS$_d$ slices
and $\partial\partial \mathcal M:=\lbrace\mathcal M \,\vert\, y=\epsilon_1, z=\epsilon_2\rbrace$
is the boundary of $\partial_1\mathcal M$,
see figure~\ref{fig:adsads-bndy}.

We briefly discuss the boundary conditions to be imposed on the various parts of the boundary.
On the $\mathbb Z_2$-fixed part at $R=0$/$y=2l$, definite orbifold parity demands either vanishing function
value $\phi=0$ or vanishing normal derivative $\partial^{}_R \phi=0$.
In view of the decomposition $\phi=\varphi f$ discussed in section~\ref{sec:KGfield},
this places restrictions on $f$.
Further restrictions are imposed on $f$ by
Dirichlet$_{d+1}$/Neumann$_{d+1}$ or more general mixed boundary conditions
at $R\rightarrow\infty$/$y=0$.
For non-Dirichlet boundary conditions and $\nu\geq 1$, the inner product needs to be properly renormalized, as usual.
On the remaining part, which is the boundary of the AdS$_d$ slices at $z=0$,
Dirichlet$_d$/Neumann$_d$ or mixed boundary conditions can be imposed.
We focus on Dirichlet$_d$ first and discuss non-Dirichlet boundary conditions in section~\ref{sec:z-Neumann}.
Finally, regularity and normalizability at the origin of the AdS$_d$ slices at $z=\pi/2$ places restrictions
on the AdS$_d$ modes $\varphi$.
This condition is satisfied by choosing for $\varphi$ the modes discussed in \cite{Balasubramanian:1998sn},
which decay as a power-law at the origin.
In the following we set $L=l=1$.

Focusing on Dirichlet$_d$ boundary conditions we now construct the counterterms that render the action
finite and stationary when both the equations of motion and the boundary conditions hold.
Using integration by parts and dropping terms which
vanish by orbifold parity and normalizability at $z=\pi/2$ the
action (\ref{eqn:KGaction}) reads
\begin{equation}
 S=\frac{1}{2}\int_{\partial_1^{}\mathcal M}\phi\sqrt{g^{yy}}\partial_y\phi
   +\frac{1}{2}\int_{\partial_2\mathcal M}\phi\sqrt{g^{zz}}\partial_z\phi+\text{EOM}~.
\end{equation}
The volume forms are suppressed throughout, they are the standard forms constructed from the (induced) metric on the respective (sub)manifold.
Note that both terms are divergent for $\epsilon_1\rightarrow 0$.
The familiar divergence of the first term
has to be cancelled by counterterms on $\partial_1\mathcal M$.
The second term is divergent due to the
integral over $y\in(\epsilon_1,2l)$.
Expanding the integrand in $y$ and performing the integral order by order
we isolate the divergent part, which is to be cancelled by
a counterterm on $\partial\partial\mathcal M$.
For $\nu\in(1,2)$ we find the counterterms
\begin{subequations}
\begin{align}\label{eqn:SctdM}
S^{}_{\partial\mathcal M}&=-\frac{1}{2}\int_{\partial^{}_1\mathcal M}
 \left[  \left(\frac{d}{2}-\nu\right)\phi^2
        +\frac{1}{2(\nu-1)}\phi\,\square^W_{g_\mathrm{ind}}\phi\right]
 +\frac{1}{4(\nu-1)}\int_{\partial\partial\mathcal M}\phi\mathcal L_{n}\phi
 ~,
\end{align}
where $\mathcal L_{n}$ is the Lie derivative along $n=-\sqrt{g^{zz}}\partial_z$,
which is the outward-pointing normalized vector field in $T\partial_1\mathcal M$ normal to $\partial\partial \mathcal M$.
$\square^W_{g_\mathrm{ind}}$ is defined below (\ref{eqn:asymptField}) and $g_\mathrm{ind}$ is the induced metric.
For $\nu\in(0,1)$ the second term in the first integral in (\ref{eqn:SctdM}) is absent,
such that the boundary terms do not contain derivatives.
For $\nu=1$ we find
\begin{align}
S^{}_{\partial\mathcal M}&=
 -\frac{1}{2}\int_{\partial^{}_1\mathcal M}
 \left[  \left(\frac{d}{2}-1\right)\phi^2
         -\big(\log y+\kappa\big)\phi\,\square^W_{g_\mathrm{ind}}\phi\right]
  -\frac{1}{2}\int_{\partial\partial\mathcal M} (\log y+\kappa)\phi\mathcal L_{n}\phi~.\label{eqn:SctdMnu1}
\end{align}
\end{subequations}
Here we have included, with an arbitrary coefficient $\kappa$,
a combination of boundary terms which is compatible with all bulk symmetries and finite for $\nu=1$.
Note that invariance under radial isometries,
corresponding to conformal transformations on the boundary, is broken by the $\log y$ terms.
We also emphasize that partial integration in the counterterms has to be carried out carefully,
since e.g.\ $\partial_1\mathcal M$ itself has a boundary.
The counterterms also enter the symplectic structure and the associated inner product. Following \cite{Compere:2008us}, we find
\begin{subequations}
\begin{align}
 \langle\phi_1,\phi_2\rangle^{}_\mathrm{ren}&=\langle\phi_1,\phi_2\rangle^{}_{\mathcal M}
-\frac{1}{2(\nu-1)}\langle\phi_1,\phi_2\rangle^{}_{\partial^{}_1\mathcal M}~,
 &\nu&\in(1,2)~,
\label{eqn:ren-prod}\\
 \langle\phi_1,\phi_2\rangle^{}_\mathrm{ren}&=\langle\phi_1,\phi_2\rangle^{}_{\mathcal M}
+\big(\log\epsilon_1+\kappa\big)\langle\phi_1,\phi_2\rangle^{}_{\partial^{}_1\mathcal M}~,
& \nu&=1~.\label{eqn:ren-prod-nu1}
\end{align}
\end{subequations}

For calculating CFT correlation functions we need the variations of the action to be
finite when evaluated on-shell.
The variation of $S_\mathrm{ren}:=S+S^{}_{\partial\mathcal M}$ reads
\begin{align}\label{eqn:deltaS1}
 \delta S_\mathrm{ren}=
  \text{EOM} & - \int_{\partial^{}_0\mathcal M}\delta\phi\sqrt{g^{yy}}\partial_y\phi
               +\int_{\partial^{}_2\mathcal M} \delta\phi\sqrt{g^{zz}}\partial_z\phi
               + \delta S_\mathrm{ren}^\nu~.
\end{align}
The first boundary term vanishes for solutions with definite $\mathbb Z_2$ parity.
The $\partial_2^{}\mathcal M$ integral is divergent for $\epsilon_1\rightarrow 0$ and the remaining part is
\begin{subequations}\label{eqn:deltaS2}
\begin{align}
   \delta S_\mathrm{ren}^{\nu} &= \int_{\partial_1^{}\mathcal M}
        2\nu\phi^\sbr{0}_\text{N}\delta\phi^\sbr{0}_\text{D}
       +\frac{1}{2(1-\nu)}\int_{\partial\partial\mathcal M}
            \delta\phi \sqrt{g^{zz}}\partial^{}_z \phi~,
\\
   \delta S_\mathrm{ren}^{\nu} &=\int_{\partial_1^{}\mathcal M}
        \delta\phi^\sbr{0}_\text{D}\big(2\phi^\sbr{0}_\text{N}+(1-2\kappa)\phi^\sbr{2}_\text{D}\big)
       +\int_{\partial\partial\mathcal M}
           (\log y+\kappa)\delta\phi \sqrt{g^{zz}}\partial^{}_z \phi  ~,
\label{eqn:deltaSdMnu1}
\end{align}
\end{subequations}
for $\nu\in(1,2)$ and $\nu=1$, respectively.
The $\partial\partial\mathcal M$ terms are divergent for $\epsilon_1\rightarrow 0$ and combine
with the divergent $\partial_2\mathcal M$ term in (\ref{eqn:deltaS1}) to render the
variation finite as we remove the cut-off on $y$.
For fixed Dirichlet$_d$ boundary conditions there are no divergences for $\epsilon_2\rightarrow 0$,
such that the limit $\epsilon_{1/2}\rightarrow 0^+$ is finite and independent of the order
in which the limits are performed.
Thus, we have renormalized the theory such that we have finite variations with respect to the boundary
data at $y=0$, while keeping fixed Dirichlet boundary conditions at $z=0$.
This allows to compute correlators for the dual CFT on AdS$_d$ with fixed boundary conditions.

In summary, the renormalized action is stationary for solutions of the Klein-Gordon equation with Dirichlet$_d$ boundary conditions,
provided they have definite $\mathbb Z_2$ parity
such that the
$\partial^{}_0\mathcal M$ integral in (\ref{eqn:deltaS1}) vanishes
and satisfy either the Dirichlet$_{d+1}$ condition $\delta \phi^\sbr{0}_\text{D}=0$
or the Neumann$_{d+1}$ condition
\begin{align} \label{eqn:NeumannBC}
 \phi^\sbr{0}_\text{N}=0~,\quad \nu\in(1,2)~,\qquad\qquad
  2\phi^\sbr{0}_\text{N}+(1-2\kappa)\phi^\sbr{2}_\text{D}=0~,\quad \nu=1~,
\end{align}
such that the $\partial_1^{}\mathcal M$ integral in (\ref{eqn:deltaS2}) vanishes.
The remaining finite combination of the $\partial_2^{}\mathcal  M$ and $\partial\partial\mathcal M$ integrals
vanishes for Dirichlet$_d$ boundary conditions. This can be seen as follows, expanding
\begin{equation}
 \varphi=z^{\frac{d-1}{2}-\mu}\big(\varphi_\text{D}^\sbr{0}+\dots\big)+
         z^{\frac{d-1}{2}+\mu}\big(\varphi_\text{N}^\sbr{0}+\dots\big)~,
\end{equation}
where $\mu$ is defined in \eqref{eqn:defmu},
and using the fixed Dirichlet$_d$ boundary condition $\varphi_\text{D}^\sbr{0} = \delta\varphi_\text{D}^\sbr{0} =  0$,
bilinears in $\phi$, $\delta\phi$ scale at least as $z^{d-1+2\mu}$.
$\sqrt{g^{zz}}\partial_z$ does not decrease the order in $z$ and
the volume forms on $\partial_2\mathcal M$, $\partial\partial\mathcal M$ are $\propto z^{-(d-1)}$.
Thus, the overall scaling is with a positive power of $z$
and as the integrations are performed for fixed $z=\epsilon_2$ the integrands vanish for $\epsilon_2\rightarrow 0$.

\subsection[Dirichlet\texorpdfstring{$_d$}{(d)} beyond the unitarity bound]{Dirichlet\texorpdfstring{\boldmath{$_d$}}{(d)} beyond the unitarity bound}
\label{sec:DirichletBeyond}

With the renormalization set up in the previous section, we now study
the bulk theory in the mass range corresponding to a CFT with an operator violating the unitarity bound.
We use the decomposition $\phi = \varphi f$ discussed in section~\ref{sec:KGfield}
and determine the spectrum from the boundary conditions at $y=0$ and $\mathbb Z_2$ parity, which
impose restrictions on the radial profiles $f$.
This yields a quantization condition on the `AdS$_d$ mass' $M$ introduced in section~\ref{sec:KGfield},
which we parametrize by a complex parameter $\mu$ as
\begin{equation}\label{eqn:defmu}
 M^2=:-\frac{(d-1)^2}{4}+\mu^2~.
\end{equation}
Note that modes with $\mu \in \mathbb{R}$ respect the AdS$_d$ BF bound.
We start with non-integer $\nu$ and discuss the case $\nu=1$ separately.
For completeness we discuss both Neumann$_{d+1}$ and Dirichlet$_{d+1}$ boundary conditions,
but of course expect unitarity violations only for the former.

The two independent solutions to the radial equation (\ref{eqn:radial-eq}) for non-integer $\nu$ are given by
\begin{equation}\label{eqn:radial-modes}
 f^{}_\mathrm{N/D}=(\cosh R)^{-\frac{d}{2}}\: P_{\mu-\frac{1}{2}}^{a^{}_\mathrm{N/D}\nu}\big(\tanh R\big)~,\qquad a^{}_\mathrm{N}=1,\quad a^{}_\mathrm{D}=-1~,
\end{equation}
where $P_\alpha^\beta$ are the generalized Legendre functions.
For the discussion of Dirichlet$_{d+1}$ and Neumann$_{d+1}$ boundary conditions we use the radial variable $y=2e^{-R}$, see section~\ref{sec:asympt}.
The asymptotic expansions of the radial modes (\ref{eqn:radial-modes}) around the conformal boundary at $y=0$ are given by
$f^{}_\mathrm{N/D}=y^{\frac{d}{2}-a^{}_\mathrm{N/D}\nu}\big(2^{a^{}_\mathrm{N/D}\nu}/\Gamma(1-a^{}_\mathrm{N/D}\nu)+\dots\big)$,
where the ellipsis denotes subleading terms of integer order.
Hence, we conclude that modes with radial profile $f_\mathrm{N}$/$f_\mathrm{D}$ satisfy
Neumann$_{d+1}$/Dirichlet$_{d+1}$ boundary conditions.
Imposing definite $\mathbb Z_2$ parity translates to the conditions
$f\vert^{}_{R=0}=0$ for odd
and $f^\prime\vert^{}_{R=0}=0$ for even parity.
For the modes (\ref{eqn:radial-eq}) we have
\begin{subequations}
\begin{flalign}
 f^{}_{\mathrm{D/N}}(0)&=
  \frac{\sqrt{\pi }\, 2^{a^{}_\mathrm{D/N} \nu}}{\Gamma \left(\frac{3}{4}-\frac{\mu }{2}-a^{}_\mathrm{D/N}\frac{\nu}{2}\right) \Gamma \left(\frac{3}{4}+\frac{\mu }{2}-a^{}_\mathrm{D/N}\frac{\nu
   }{2}\right)}~,\label{eqn:funcatorigin}\\
 f_{\mathrm{D/N}}^\prime(0)&=
    \frac{-\sqrt{\pi }\, 2^{1+a^{}_\mathrm{D/N}\nu}}{\Gamma \left(\frac{1}{4}-\frac{\mu }{2}-a^{}_\mathrm{D/N}\frac{\nu }{2}\right) \Gamma \left(\frac{1}{4}+\frac{\mu }{2}-a^{}_\mathrm{D/N}\frac{\nu
   }{2}\right)}~.\label{eqn:derivatorigin}
\end{flalign}
\end{subequations}
The expressions on the right hand sides vanish when the appropriate $\Gamma$-functions in the denominator have a pole,
which is for non-positive integer arguments. The spectrum can therefore be read off from
(\ref{eqn:funcatorigin}) for odd and  (\ref{eqn:derivatorigin}) for even $\mathbb Z_2$ parity, which yields
\begin{equation} \label{eqn:spectrum}
 \mu_\mathrm{D/N, even/odd}^2
   =\Big(2n+\frac{1}{2}-a^{}_\mathrm{D/N}\nu+b^{}_{\mathrm{even/odd}}\Big)^2~,\qquad n\in\mathbb N~,
\end{equation}
where $b^{}_\mathrm{even}=0$ and $b^{}_\text{odd}=1$ for even and odd parity, respectively.
Note that these $\mu$ are real, such that the transverse modes $\varphi$ of the bulk field with Dirichlet$_{d+1}$/Neumann$_{d+1}$
boundary conditions do not violate the AdS$_d$ BF bound.
However, there can be modes with $\mu=0$ which saturate the BF bound for half-integer $\nu$ and Neumann$_{d+1}$ boundary condition.

For a concrete realization of the transverse modes $\varphi$ we use the AdS modes discussed in \cite{Balasubramanian:1998sn}.
Imposing normalizability at the origin and boundary conditions on the conformal boundary of the AdS$_d$ slices yields
a quantization of their frequencies depending on $\mu$. For the Dirichlet$_d$ case, all the modes are normalizable with respect to the usual symplectic structure and the frequencies $\omega$ are given by
\begin{equation}\label{eqn:freqDads}
\omega^{}_\mathrm{D/N, even/odd}= \pm \left[ \ell+ 2 p + \frac{d-1}{2}+ \mu^{}_\mathrm{D/N, even/odd}\right]~,
\qquad p \in\mathbb N~,
\end{equation}
where  $\ell$ denotes the principal angular momentum.
Note that the subscripts $\mathrm{D/N}$ in (\ref{eqn:freqDads}) refer to Dirichlet$_{d+1}$/Neumann$_{d+1}$ boundary conditions on the conformal boundary of AdS$_{d+1}$.
For the case of Neumann$_d$ to be discussed in section~\ref{sec:z-Neumann}, the frequencies are given by (\ref{eqn:freqDads})
with ${\mu^{}_\mathrm{D/N, even/odd}\,{\rightarrow}\, -\mu^{}_\mathrm{D/N, even/odd}}$.

For Dirichlet$_d$ boundary conditions the AdS$_d$ norms are positive \cite{Andrade:2011dg}, so the existence of ghosts
depends only on the norms of the radial modes, which we now calculate.
With the decomposition $\phi = \varphi(x) f(R)$ the renormalized inner product
(\ref{eqn:ren-prod}) reads
$\langle \phi_1 , \phi_2 \rangle^{}_{{\cal M}} = \langle \varphi_1 , \varphi_2 \rangle^{}_\mathrm{slice}  \langle f_1 , f_2 \rangle^{}_\mathrm{SL, ren}$~,
where the renormalized SL product is given by
\begin{equation}\label{SL renorm}
    \langle f_1 , f_2 \rangle^{}_\mathrm{SL, ren} = \langle f_1 , f_2 \rangle^{}_\mathrm{SL} - \frac{1}{2(\nu-1)}(\cosh R)^{d-2} f_1^\ast f^{}_2 \big|_{R \rightarrow \infty}~.
\end{equation}
We evaluate (\ref{SL renorm}) using (\ref{eqn:reducedSL}) and the modes $f_N$/$f_D$ given in (\ref{eqn:radial-modes}) which satisfy
Neumann$_{d+1}$/Dirichlet$_{d+1}$ boundary conditions for all $\mu$.
The divergence of the bare SL product $\langle f_1 , f_2 \rangle_{SL}$ for Neumann$_{d+1}$ and $\nu\in(1,2)$ is cancelled by the boundary term,
such that the inner product is finite.
Furthermore, the finite contribution from $R=\infty$ vanishes for all $\mu$ if Neumann$_{d+1}$/Dirichlet$_{d+1}$ boundary conditions are satisfied.
The inner product thus evaluates to
\begin{equation}\label{eqn:SLadsfin}
    \langle f_1 , f_2 \rangle^{}_\mathrm{SL, ren} =  - \frac{1}{M_1^2 - M_2^2} (\cosh R)^d ( f_1^\ast f^\prime_2 - {f^\prime_1}^\ast f^{}_2  ) \big|_{R=0}~.
\end{equation}
Note that the term in parenthesis on the right hand side vanishes if orbifold boundary conditions are satisfied,
and therefore modes with different boundary mass are orthogonal.
The expression (\ref{eqn:SLadsfin}) as it stands is not defined for $M_1=M_2$.
However, it can be extended continuously to coinciding masses given by (\ref{eqn:spectrum}),
since in that case the term in parenthesis vanishes as well.
Defining $||f||^2 := \langle f, f \rangle^{}_\mathrm{SL, ren}$, the inner product for AdS$_{d+1}$ fields $\phi_\mathrm{D/N}$ with Dirichlet$_{d+1}$/Neumann$_{d+1}$ boundary conditions reads
\begin{equation}\label{eqn:innerproductAdS}
  \langle \phi^{}_{\mathrm{D,1}} , \phi^{}_{\mathrm{D,2}} \rangle^{}_{{\cal M}} = \delta^{}_{M_1 M_2} \langle \varphi^{}_1 , \varphi^{}_2 \rangle^{}_\mathrm{slice}\, || f^{}_\mathrm{D} ||^2~,
   \qquad
  \langle \phi^{}_{\mathrm{N,1}} , \phi^{}_{\mathrm{N,2}} \rangle^{}_{{\cal M}} = \delta^{}_{M_1 M_2} \langle \varphi^{}_1 , \varphi^{}_2 \rangle^{}_\mathrm{slice}\, || f^{}_\mathrm{N} ||^2~.
\end{equation}

For the Dirichlet$_{d+1}$ modes with even/odd $\mathbb Z_2$ parity we find
\begin{flalign}\label{eqn:normAdS_D}
 || f_\mathrm{D,even/odd} ||^2 &=\frac{(2 n+b_\mathrm{even/odd})!}{(1+2 \nu+4 n +2b_\mathrm{even/odd}) \Gamma (1+2 \nu +2 n+b_\mathrm{even/odd})}~.
\end{flalign}
As expected, these are positive for all $n\in\mathbb N$ and $\nu\geq 0$.
Thus, since $<\cdot,\cdot>^{}_\mathrm{slice}$ is non-negative for Dirichlet$_d$ boundary conditions,
the spectrum is ghost-free.
Similarly, for Neumann$_{d+1}$ boundary conditions we find the norms
\begin{flalign}\label{eqn:normAdS_N}
 || f_\mathrm{N,even/odd} ||^2 &=\frac{(2 n+b_\mathrm{even/odd})!}{(1-2\nu+4 n+2b_\mathrm{even/odd}) \Gamma (1-2 \nu +2 n+b_\mathrm{even/odd})}~,
\end{flalign}
which are positive for $\nu\in[0,1)$ as expected.
For $\nu>1$ we first consider $\nu\notin\mathbb Z+\frac{1}{2}$.
If $m:=\lfloor 2\nu-b_\mathrm{even/odd}\rfloor$, where $\lfloor\cdot\rfloor$ denotes the integer part,
is even, the $n=0$ mode has negative norm since the coefficient of the $\Gamma$-function in the denominator is
negative while the $\Gamma$-function itself is positive.
If $m$ is odd the $n=(m-1)/2$ mode is of negative norm since the coefficient is positive while the $\Gamma$-function is negative.
As $<\cdot,\cdot>^{}_\mathrm{slice}$ is non-negative we thus
have ghosts in the spectrum for non-half-integer $\nu$ in $(1,2)$,
such that the non-unitarity of the dual
boundary theory is nicely reflected in the bulk.
For $\nu=k+\frac{1}{2}$, $1\leq k\in\mathbb Z$ the AdS$_d$ modes have integer $\mu$ and by continuity of (\ref{eqn:normAdS_N}) there are
modes with vanishing or negative norm.
The $n=0$ and $n=k-b_\mathrm{even/odd}$ modes are of norm zero and yield the same $\mu^2$ unless $k=1$
with odd $\mathbb Z_2$ parity\footnote{For $k$ large enough there are further pairs of modes with norm zero and the same $\mu^2$,
which are of the form $(n_1,n_2)=(1,n-b_\mathrm{even/odd}-1)\,,\, (2,n-b_\mathrm{even/odd}-2)$ etc.}.
This degeneracy in the spectrum indicates that the basis of solutions we are using is incomplete,
so we expect `logarithmic modes', analogous to those in \cite{Grumiller:2008qz}.
Once the log-modes are incorporated, continuity of the spectrum indicates that ghosts must be present \cite{Andrade:2009ae}.
For $k=1$ with odd parity the $n=0$ mode is of negative norm and the others are positive,
such that the non-unitarity of the dual theory is reproduced in the bulk.

We close this section noting that the results established explicitly here for $\nu<2$ extend to higher $\nu\notin\mathbb Z$
even without knowledge of the exact counterterms.
We consider the expansions of the Dirichlet and Neumann bulk fields near the conformal boundary $\partial_1 {\cal M}$
\begin{equation}\label{}
  \phi^{}_\mathrm{D} = y^{\frac{d}{2}-\nu} \sum_{k} \phi_\mathrm{D}^\sbr{2k} y^{2k}~,
   \qquad
 \phi^{}_\mathrm{N} = y^{\frac{d}{2}+\nu} \sum_{k} \phi_\mathrm{N}^\sbr{2k} y^{2k}~,
\end{equation}
where $k$ is a non-negative integer.
For $2\nu\notin\mathbb Z$ the only way to get boundary terms which are quadratic in the field
and have an integer scaling -- finite terms, in particular --
is through the combination $\phi^{}_\mathrm{D} \phi^{}_\mathrm{N}$.
However, since $\phi^{}_\mathrm{D} \phi^{}_\mathrm{N}=\mathcal O(y^d)$ and derivative/curvature-terms are subleading
with even powers of $y$,
while the volume form is $\mathcal O(y^{-d})$, only the boundary term without derivatives can yield a finite part.
This implies that there are no extra finite contributions to the norm from the additional boundary terms.
For half-integer $\nu$ the combination of two $\phi^{}_\mathrm{D}$ fields with the volume form scales as an odd power of $y$,
so such terms can again not yield finite contributions.
The results (\ref{eqn:normAdS_D}), (\ref{eqn:normAdS_N}) are therefore also valid for generic $\nu\notin\mathbb Z$.

\subsubsection{Saturating the unitarity bound}\label{sec:saturateUnitarityAdS}
We now consider the special case $\nu = 1$, corresponding on the boundary to an operator which saturates the
unitarity bound\footnote{\label{ftnt} Here we slightly abuse notation --
boundary conformal symmetry is broken for integer $\nu$ due to the logarithmic counterterms, such that the unitarity
bound does not strictly apply.}.
The modes (\ref{eqn:radial-modes}) are not linearly independent for integer $\nu$, so we instead use the basis of radial profiles
\begin{align}\label{eqn:radial-modes-nu1}
  f^{}_i(R)&=u^{2c_i-\frac{3}{2}}\left(1-u^2\right)^{\frac{d+2}{4}} \, _2F_1\Big(c_i-\frac{\mu}{2},c_i+\frac{\mu}{2};2c_i-1;u^2\Big)~,\qquad i=1,2~,
\end{align}
where $u=\tanh(R)$ and $c_1=3/4$, $c_2=5/4$.
Since
$ f_1(0)=1$, $ f_1^\prime(0)=0$ and
$ f_2(0)=0$, $ f_2^\prime(0)=1$,
the modes are independent and $f_1$/$f_2$ has even/odd $\mathbb Z_2$ parity.
The expansions around $y=0$ are
\begin{equation}\label{eqn:nu1asympt}
  f_i=\sqrt{\pi }\, 2^{1-i} y^{\frac{d}{2}-1} \big( f_i^{(0)}+y^2\log(y) f_i^{(1)}+y^2 f_i^{(2)}+\dots\big)~,
\end{equation}
where
$f_i^{(0)}=1/\big(\Gamma \left(c_i-\frac{\mu }{2}\right) \Gamma \left(c_i+\frac{\mu }{2}\right)\big)$\:,\:
$f_i^{(1)}=\frac{1}{8}(1-4\mu^2)f_i^{(0)}$ and
\begin{flalign}
 f_i^{(2)}&=
\frac{\left(1-4 \mu ^2\right) \left(\psi(c_i-\frac{\mu }{2})+\psi(c_i+\frac{\mu }{2})+2\gamma-1\right)-2d-(-1)^i4}{
 16\Gamma\left(c_i-\frac{\mu }{2}\right) \Gamma \left(c_i+\frac{\mu }{2}\right)}~.\label{eqn:nu1NeumannCoeff}
\end{flalign}
Here, $\gamma$ is the Euler-Mascheroni constant and $\psi$ the digamma function.

We now discuss the spectrum, which for even/odd $\mathbb Z_2$ parity is found by imposing Dirichlet$_{d+1}$ or Neumann$_{d+1}$
boundary conditions on $f_1$/$f_2$.
We first consider Dirichlet boundary conditions, which amount to setting to zero the leading coefficient
in (\ref{eqn:nu1asympt}), i.e. $f_i^{(0)} = 0$.
This yields $\mu^{}_{i,\mathrm{D}}=\pm 2(n+c_i)$ with $n\in\mathbb N$.
Note that for these choices of $\mu$ the coefficients $f_i^{(2)}$ are finite despite the pole in the
denominator, namely  ${f_i^{(2)}\big\vert^{}_{\mu_{i,\mathrm{D}}}=(-1)^n(n+2)!/\Gamma(n-1+2c_i)}$.
This ensures that the modes are non-trivial.
The norms for the Dirichlet case are positive $\forall n\in\mathbb N$, as expected:
\begin{equation}
 || f^{}_{i,\mathrm{D}}||^2  =\pi n!(n+1)!\frac{n+2c_i-1}{4^i(n+c_i)\Gamma(n+2c_i)^2}~.
\end{equation}

We now come to the Neumann$_{d+1}$ boundary condition which, as seen in (\ref{eqn:NeumannBC}), amounts to
\begin{equation}\label{eqn:nu1NeumannCond}
 2 f_i^{(2)}+(1-2\kappa) f_i^{(1)}=0~.
\end{equation}
The specific solution $\mu^2=\frac{1}{4}$ only exists for $d=2$ and even parity.
For the remaining solutions we note that,
as $f_i^{(2)}$ is finite for the choices of $\mu$ which make its denominator diverge
while $f_i^{(1)}$ vanishes, those $\mu$ do
not yield solutions.
Therefore, (\ref{eqn:nu1NeumannCond}) is equivalent to
\begin{equation}\label{eqn:nu1NeumannSpectrum}
 \psi(c_i-\frac{\mu}{2})+\psi(c_i+\frac{\mu}{2})+2(\gamma-\kappa)=\frac{2d+(-1)^i4}{1-4\mu^2}~.
\end{equation}
We first argue that there are only real or purely imaginary solutions.
Assume that we have $\mu=a+i b$ with $a,b\neq 0$ satisfying (\ref{eqn:nu1NeumannSpectrum}).
Due to the non-vanishing imaginary part of $\mu$ the arguments of the digamma functions
in (\ref{eqn:nu1NeumannSpectrum}) are non-integer,
such that we can expand
$\psi(1+z)+\gamma=\sum_{n=1}^\infty\frac{z}{n(n+z)}$ \cite{abramowitz+stegun}.
Taking the imaginary part of (\ref{eqn:nu1NeumannSpectrum}) then yields
\begin{equation} \label{eqn:nu1NeumannSpectrum4}
-\sum_{n=0}^\infty \frac{n+c_i}{\big |n+c_i+\mu/2\big |^2\,\big |n+c_i-\mu/2\big |^2}
=16\,\frac{d+(-1)^i2}{|1-4\mu^2|^2}~.
\end{equation}
Since each term in the sum on the left hand side is positive,
the overall left hand side is negative. On the other hand, the right hand side is non-negative, which yields a contradiction.
The assumption that there are solutions with $a,b\neq 0$ therefore has to be dropped
and we only have real or purely imaginary solutions. For such $\mu$'s the modes (\ref{eqn:radial-modes-nu1}) are real.

We now focus on the lowest-$M^2$ solutions.
To see whether we have tachyons, i.e.\ states with negative mass squared below the BF bound,
we consider purely imaginary $\mu=i \lambda$.
Equation (\ref{eqn:nu1NeumannSpectrum}) then becomes
\begin{equation}
 \Re \psi(c_i+\frac{i\lambda}{2})+\gamma-\kappa=\frac{d+(-1)^i2}{1+4\lambda^2}~.
\end{equation}
Both sides of the equation are monotonic functions of $|\lambda|$.
While the right hand side decreases with $|\lambda|$ and tends to zero,
the left hand side increases and tends to infinity.
Thus, we have one tachyonic state if at $\lambda=0$ the
right hand side is greater than the left hand side, and none otherwise.
This translates to $\kappa>2-d-\log 8-(-1)^i\frac{\pi}{2}$ as condition
for the existence of a tachyon in the spectrum.
A discussion of such a tachyonic scalar field with negative mass squared below the BF bound in global AdS is given in Appendix \ref{app:tachyon-globalAdS}.
If $\kappa$ is such that there is no tachyon, the left hand side of (\ref{eqn:nu1NeumannSpectrum}) is greater than or equal to
the right hand side at $\mu=0$.
Since the left hand side is bounded for real $\mu\in[0,\frac{1}{2})$, while the right hand side tends to $+\infty$
for $\mu\rightarrow \frac{1}{2}$,
the lowest-$M^2$ solution in that case has $\mu\in[0,\frac{1}{2})$.

With the properties of the spectrum discussed above we can now examine the norms.
Calculating the renormalized inner product for the modes (\ref{eqn:radial-modes-nu1})
and using (\ref{eqn:nu1NeumannSpectrum}) to simplify it, we find
\begin{equation}
 ||f^{}_{i,\mathrm{N}}||^2=\pi
       \frac{-8\frac{d+(-1)^i2}{1-4\mu^2}
        +\frac{1-4\mu^2}{4\mu}\big(\psi^\sbr{1}(c_i+\frac{\mu}{2})-\psi^\sbr{1}(c_i-\frac{\mu}{2})\big)}{2^{4c_i}\Gamma(c_i-\frac{\mu}{2})^2\,\Gamma(c_i+\frac{\mu}{2})^2}~.
\end{equation}
where $\psi^\sbr{1}$ is the trigamma function.
The denominator is positive for $\mu$ real or purely imaginary, so the sign of the norm only depends on the numerator.
For any choice of $\kappa$, $d\geq 2$ and
$\mu\in[0,\frac{1}{2})$ or $\mu$ purely imaginary,
the first term in the numerator is non-positive and
the second one negative, such that the norm is negative and we find ghosts in all these cases.

\subsection[Neumann\texorpdfstring{$_d$}{(d)} and dimensional reduction]{Neumann\texorpdfstring{\boldmath{$_d$}}{(d)} and dimensional reduction} \label{sec:z-Neumann}
In this section we study the case of Neumann$_d$ boundary conditions at the boundary of the AdS$_d$ slices
for non-integer $\nu$.
It turns out that normalizability in that case is quite delicate,
as we will see shortly. We expect similar features for all boundary conditions which allow the Neumann$_d$ modes to fluctuate.
The values of $\mu^2$ corresponding to Neumann$_{d+1}$/Dirichlet$_{d+1}$ and even/odd orbifold parity
were given in  (\ref{eqn:spectrum}) and read
\begin{equation}\label{eqn:spectrum2}
 \mu_{\mathrm{D/N, even/odd}}^2
   =\Big(2n+\frac{1}{2}-a^{}_\mathrm{D/N}\nu+b^{}_{\text{even}/\text{odd}}\Big)^2~,\qquad n\in\mathbb N~.
\end{equation}
In the previous section we have discussed Dirichlet$_d$ boundary conditions.
In that case, normalizability of the inner product (\ref{eqn:inner-prod-factorized}) was equivalent to
normalizability of the radial part $\langle f^{}_1,f^{}_2\rangle^{}_\mathrm{SL}$ since the transverse part
$\langle\varphi^{}_1,\varphi^{}_2\rangle^{}_\mathrm{slice}$ was finite.
On the other hand, for Neumann$_d$ normalizability of $\langle\varphi^{}_1,\varphi^{}_2\rangle^{}_\mathrm{slice}$
is not given a priori.
Instead, counterterms on $\partial^{}_2\mathcal M$ with their contribution to the symplectic structure
are needed to render the associated inner product 
$\langle\varphi^{}_1,\varphi^{}_2\rangle^{}_\mathrm{slice, ren}$ finite for $\mu^2> 1$.
Additional terms on $\partial\partial\mathcal M$ may also be required to cancel combined divergences in the radial
and slice parts of the inner product.
The (standard) geometric counterterm action on $\partial_2 \mathcal M$ can be arranged as a series of terms with decreasing
degree of divergence as $z\rightarrow 0$
\begin{equation} \label{eqn:AdSNeumann-d-ct}
 S^{}_{\partial^{}_2\mathcal M}=\int_{\partial^{}_2\mathcal M} \alpha\, \phi^2+\beta\, \phi\square_{g_\mathrm{ind}}^{}\phi+\gamma\, \mathcal R[g_\mathrm{ind}] \phi^2+...~,
\end{equation}
with fixed coefficients $\alpha$, $\beta$ etc.
Note that, to render modes with $\mu^2>1$ normalizable, the precise relation between the coefficients and $\mu$ would be required,
as seen for the corresponding relation to $\nu$ in \eqref{eqn:SctdM}. 
Thus, with fixed coefficients we could at most render the modes for one $\mu^2>1$ normalizable\footnote{
We have not attempted to generate coefficients like $1/(1-\mu)$ by acting on $\phi$ with non-local operators
containing e.g.\ inverse $z$-derivatives.
}.
Moreover, calculating explicitly the contribution to the inner product 
and factorizing off the radial 
part similarly to \eqref{eqn:inner-prod-factorized},
we find
$\langle \phi^{}_1,\phi^{}_2\rangle^{}_{\partial_2\mathcal M}=
 \langle \varphi^{}_1,\varphi^{}_2\rangle^{}_{} \int_0^\infty dR\, \lambda^{d-3}f_1^\ast f^{}_2$~,
where $\langle \varphi^{}_1,\varphi^{}_2\rangle^{}_{}$ denotes the $R$-independent part at fixed $z$.
Note that the radial part is not the Sturm-Liouville inner product
and therefore -- in contrast to the counterterm contribution from $\partial_1\mathcal M$ -- the renormalization
can not be absorbed by renormalizing only one of the factors in \eqref{eqn:inner-prod-factorized}.
Thus, the counterterms can not cancel the divergences coming solely
from the $\langle .,.\rangle^{}_\mathrm{slice}$ part of the inner 
product \eqref{eqn:inner-prod-factorized} and only the modes with $\mu^2<1$ are normalizable.
Finally, we note that since the boundary geometry is global AdS$_d$ the
results of \cite{Andrade:2011dg} apply and we conclude that even if the AdS$_d$ part 
$\langle\varphi^{}_1,\varphi^{}_2\rangle^{}_\mathrm{slice}$
of \eqref{eqn:inner-prod-factorized}
was properly renormalized, 
it would be indefinite for $\mu^2>1$.
The bulk theory would then contain ghosts immediately and we 
therefore choose to add no counterterms on $\partial_2\mathcal M$.

In summary, the Neumann$_d$ theory is specified by the AdS$_{d+1}$ action (\ref{eqn:KGaction})
with mass parametrized by $\nu$, and the $\nu$-dependent counterterms discussed in
section~\ref{sec:BCandRenormalization} to render the inner product finite as $\epsilon_1\rightarrow 0$.
The spectrum of normalizable excitations is given
by the modes with $\mu^2$ as in (\ref{eqn:spectrum2})
subject to the condition that the AdS$_d$ part of the inner product
$\langle\varphi^{}_1,\varphi^{}_2\rangle^{}_\mathrm{slice, ren}$
is finite.
This is only the case for $\mu^2<1$, i.e.
\begin{equation}
 -\frac{3}{2} < 2n-a^{}_\mathrm{D/N}\nu+b^{}_{\text{even}/\text{odd}} < \frac{1}{2}~.
\end{equation}
This immediately implies that there is at most one normalizable radial mode.
For Dirichlet$_{d+1}$ boundary conditions there are only normalizable modes for even parity and $\nu\,{\in}\,[0,\frac{1}{2})$.
They have $n=0$ and, as seen in (\ref{eqn:normAdS_D}), their norm is positive, such that there are no ghosts for Dirichlet$_{d+1}$ as expected.
The spectrum is slightly richer for Neumann$_{d+1}$.
For even $\mathbb Z_2$ parity there is a normalizable radial mode for all $\nu\geq 0$, namely $n=\lfloor (2\nu+1)/4 \rfloor$.
For odd $\mathbb Z_2$ parity we find a normalizable mode only for $\nu>1/2$, and it has $n=\lfloor (2\nu-1)/4\rfloor$.
In both of the Neumann$_{d+1}$ cases the normalizable modes -- if any -- have positive norm for
$\nu<1$ and negative norm for $\nu\,{\in}\,(1,2)$, see (\ref{eqn:normAdS_N}).
The spectrum, although somewhat trivial, is therefore ghost-free for $\nu\in(0,1)$ and contains ghosts
for $\nu\in(1,2)$, which matches our expectations based on boundary unitarity.
The discussion extends to $\nu>2$ for the same reason as discussed in section~\ref{sec:DirichletBeyond}.
Interestingly, for even $\mathbb Z_2$ parity and $\nu\,{\in}\,(2,\frac{5}{2})$, we find that the only normalizable modes have $n=1$
and positive norm, see (\ref{eqn:normAdS_N}).
Thus, the spectrum, although very simple, is free of ghosts in that case.
A similar mechanism applies to odd $\mathbb Z_2$ parity and $\nu\,{\in}\,(\frac{5}{2},3)$.

What we see here is a kind of dimensional reduction -- the radial dependence of the AdS$_{d+1}$ field $\phi$ is completely
fixed, such that it
has only  the degrees of freedom of $\varphi$, i.e.\ of the AdS$_d$ field.
The space of normalizable solutions to the Klein-Gordon equation on AdS$_{d+1}$ with mass $m^2=-d^2/4+\nu^2$ and Neumann$_d$ boundary condition is
thus isomorphic to the space of solutions to the Klein-Gordon equation on AdS$_d$ with Neumann$_d$ boundary condition and mass $M^2=-(d-1)^2/4+\mu^2$,
where $\mu^2$ is given by the only one normalizable $\mu^2<1$ of (\ref{eqn:spectrum2}).
In that sense, the bulk theory is equivalent to a boundary theory in a trivial way.

Some further comments are in order.
We notice that the bulk theory is non-local.
One way to see this is to note that with only the solutions for one $\mu^2$ available
it is impossible to localize initial data along the radial direction.
Furthermore, we note that the boundary theory lacks conformal invariance, as it is simply a
scalar field with a fixed mass, which fails to be conformally
coupled\footnote{The field equation for a conformally coupled scalar in $d$ dimensions is
$\Box \phi -\frac{1}{4}\frac{d-2}{d-1}\,\mathcal R \phi=0$, where in our case $\mathcal R  = - d(d-1)$.
This corresponds to $\mu^2=\frac{1}{4}$, which is only possible for integer $\nu$ as can be seen in \eqref{eqn:spectrum2}.
Although we have not discussed the integer-$\nu$ cases in detail for Neumann$_d$,
we expect tachyons/ghosts similar to the situation in section \ref{sec:saturateUnitarityAdS}.
}.
Hence, we do not expect the unitarity bound to hold, which explains the presence of the positive norm modes
found above.

\section{dS on the boundary}  \label{sec:dS-on-boundary}

In the present section we study the Klein-Gordon theory defined in AdS$_{d+1}$ foliated by dS$_d$ slices, which, as noted in the
introduction corresponds to a boundary dual theory defined on dS$_d$. More precisely, we take the metric to be (\ref{eqn:ads-ds}),
and set $L=H=1$ henceforth.
As mentioned in the introduction, this bulk set up is closely related to the global AdS$_{d+1}$ case discussed in \cite{Andrade:2011dg}
and  one could in principle argue that the results should translate from those in the global case, at least in conformal invariant scenarios.
However, as the dS$_d$ slicing covers only a patch of AdS$_{d+1}$ and a horizon is present,
one could expect the bulk manifestations of the unitarity violations in the boundary theory to resemble those in Poincar\'e AdS.
Furthermore, there are cases of interest in which conformal invariance is broken, and this motivates us to present our study of
the dS$_d$ case in detail. We shall also see that the spectrum possesses an interesting structure, which also makes this discussion worthwhile.

Our main interest is to find possible violations of
unitarity in the bulk when the dual theory contains an operator whose dimension violates the unitarity bound.
Thus, we will focus on Neumann boundary conditions with mass  $m^2 = -d^2/4 + \nu^2$ and $0 < \nu < 2$.
For comparison, we shall also include the Dirichlet results.
The boundary of the patch covered by the dS$_d$ foliation
consists of the causal horizon located in the interior,
where $R$ goes to zero in the coordinate system (\ref{eqn:ads-global-metric}), (\ref{eqn:coord-transf-dS}), and a piece of the
conformal boundary where $R$ goes to infinity. Below, we shall impose normalizability on the causal horizon and shall not add
counterterms in this region, in analogy to the usual treatment of the Poincar\'e horizon in the Poincar\'e patch of AdS,
see e.g. \cite{Compere:2008us}, \cite{Andrade:2011dg}, \cite{Carlip:2008jk}.
On the other hand, on the conformal boundary we will require the usual Dirichlet or Neumann boundary conditions, which can be
implemented by adding the familiar counterterms.

In order to solve the wave equation we will use the mode decomposition discussed in section~\ref{sec:KGfield}
with the dS$_d$ harmonics $\varphi=Y_{\sigma, \vec{j}}$.
Since these will play an important role in our analysis, we shall now review
their main properties closely following \cite{Marolf:2008hg}. We refer the reader to \cite{Higuchi1987} for a more extensive discussion. By definition, the dS$_d$ harmonics satisfy eigenvalue
equation
\begin{equation}\label{dS harmonics}
    \square_\gamma Y_{\sigma, \vec{j}} = - \sigma(\sigma + d - 1) Y_{\sigma, \vec{j}}~,
\end{equation}
where $\sigma$ is an arbitrary complex parameter. The collection $\vec{j}$ corresponds to $(d-1)$ angular momentum
quantum numbers, i.e. the components of $\vec{j}$ are non-negative integers such that $j_{d-1} > j_{d-2} > \ldots |j_1|$.
Note that, as a consequence of the definition (\ref{dS harmonics}), the dS harmonics are unchanged under the
replacement $\sigma \rightarrow -(\sigma + d - 1)$. Thus, without loss of generality, we can restrict ourselves
to $\Re \sigma > -(d-1)/2$ and we shall do so below.

The space spanned by the dS harmonics is endowed with the inner product,
\begin{equation}\label{dS ip}
  \langle Y_{\sigma, \vec{j}}, Y_{\sigma, \vec{k}} \rangle^{}_\mathrm{slice}  = - i \int_{\partial \Sigma} d \Omega \sqrt{g^{}_{\partial \Sigma}}\, n^i Y_{\sigma, \vec{j}}^* \stackrel {\leftrightarrow} \partial_i Y_{\sigma, \vec{k}}~.
\end{equation}
With the convention $\Re \sigma > -(d-1)/2$, it can be shown that the dS harmonics furnish unitary
representations, i.e. that (\ref{dS ip}) is positive definite, if $\sigma$ belongs to one of the following
\begin{itemize}
  \item Principal series: $\sigma = - \frac{d-1}{2} + i \rho$, with $\rho \in \mathbb{R}$,
  \item Complementary series: $- \frac{d-1}{2} < \sigma < 0$, with $\sigma \in \mathbb{R}$.
\end{itemize}
Some comments are in order here. First, it is important to keep in mind that, since we are interested in searching
for ghosts/violations of unitarity, we must consider all $\sigma$'s allowed by normalizability, and not restrict
ourselves to modes in the principal or complementary series. Second, we have defined ghosts as solutions with positive
frequency and negative norm. The notion of positive frequency we shall adopt here is closely analogous to that
depicted in \cite{Marolf:2006bk} in the context of asymptotically flat spaces foliated by dS slices.
That is, we shall choose $Y_{\sigma, \vec{j}}$ such that $\phi=Y_{\sigma, \vec{j}} f^{}_\sigma $ is positive frequency in the usual
sense in AdS$_{d+1}$. Third, we note that $\sigma_1$, $\sigma_2$ in the principal series are indistinguishable
if $\sigma_1 = \sigma_2^*$. We shall remove this ambiguity by taking $\rho > 0$ below.

\subsection{Renormalization}
\label{ren symp prod dS}

The goal of this section is to find a properly renormalized action and symplectic product for the case of the dS$_d$ slicing.
As mentioned above, we will only add to the action counterterms on the conformal boundary of the patch of AdS$_{d+1}$.
Having found a satisfactory (i.e. finite and stationary) action, we shall follow the
prescription of \cite{Compere:2008us} to determine the renormalized symplectic structure.
In analogy to the AdS$_d$ case, the action we consider is
$S^\mathrm{dS}_\mathrm{ren} = S + S^{}_{\partial \mathcal M}$,
where $S$ is given by (\ref{eqn:KGaction}) and
\begin{subequations}
\begin{align}\label{eqn:SctdM dS}
S^{}_{\partial \mathcal M} &=-\frac{1}{2}\int_{\partial \mathcal M}  \left[  \left(\frac{d}{2}-\nu \right)\phi^2
        +\frac{1}{2(\nu-1)}\phi \square^\mathrm{W}_{g_\mathrm{ind}}\phi\right] & {\rm for }& \ \nu \neq 1~,
\\ \label{action dS n=1}
   S^{}_{\partial \mathcal M}  &= -\frac{1}{2} \int_{\partial M} \left[ \left(\frac{d}{2}-1\right)\phi^2 - (\log y + \kappa) \phi \square^\mathrm{W}_{g_\mathrm{ind}} \phi \right]
   & {\rm for }& \ \nu = 1~.
\end{align}
\end{subequations}
Here $\square^\mathrm{W}_{g_\mathrm{ind}}$ is the differential operator defined in section~\ref{sec:asympt},
the radial variable $y$ is defined via $y = 2 e^{-R}$ and
$\partial {\cal M}$ denotes the part of the boundary at the radial cut-off $y=\epsilon$, see figure \ref{fig:dsads-bndy}.
Note that in (\ref{action dS n=1}) we have introduced an extra finite counterterm with an arbitrary coefficient $\kappa$.
Using the results of section~\ref{sec:asympt}, it is not hard to verify that $S^\mathrm{dS}_\mathrm{ren}$
provides a well-defined variational
principle for the relevant boundary conditions. In fact, taking an arbitrary on-shell variation we obtain
\begin{subequations}
\begin{align}\label{dS dSitter}
    \delta S^\mathrm{dS}_\mathrm{ren} &= - 2 \nu \int_{\partial  {\cal M}} \phi^\sbr{0}_\mathrm{N} \delta \phi^\sbr{0}_\mathrm{D} & {\rm for }& \ \nu \neq 1~,
\\ \label{dS dSitter nu=1}
    \delta S^\mathrm{dS}_\mathrm{ren} &= - \int_{\partial  {\cal M}} \big( 2 \phi^\sbr{0}_\mathrm{N} + (1-2\kappa) \phi^\sbr{2}_\mathrm{D} \big) \delta \phi^\sbr{0}_\mathrm{D} & {\rm for }& \ \nu = 1~,
\end{align}
\end{subequations}
where the coefficients of the asymptotic expansion are those given in section~\ref{sec:asympt}. For $0<\nu<2$, with $\nu \neq 1$,
we observe from (\ref{dS dSitter}) that $\delta S^\mathrm{dS}_\mathrm{ren}$ is indeed finite and stationary
for either Dirichlet boundary conditions, $\phi^\sbr{0}_\mathrm{D}=0$, for all $\nu$ or Neumann boundary conditions, $\phi^\sbr{0}_\mathrm{N} =0$.
In the $\nu=1$ case, (\ref{dS dSitter nu=1}) reveals that $\delta S^\mathrm{dS}_\mathrm{ren}$ is finite and stationary
for the Dirichlet boundary condition $\phi^\sbr{0}_\mathrm{D} =0$ and for
\begin{equation}\label{Neu nu=1}
    2 \phi^\sbr{0}_\mathrm{N} + (1-2\kappa) \phi_\mathrm{D}^\sbr{2} = 0~,
\end{equation}
which we shall refer to as Neumann.
The renormalized inner products constructed along the lines of \cite{Compere:2008us} read
\begin{subequations}
\begin{align}\label{ren ip}
    \langle \phi_1, \phi_2 \rangle^{}_\mathrm{ren} &= \langle \phi_1, \phi_2 \rangle^{}_{{\cal M}} - \frac{1}{2(\nu-1)} \langle \phi_1, \phi_2 \rangle^{}_{\partial {\cal M}} &{\rm for }& \ \nu \neq 1~,
\\ \label{ren ip nu=1}
    \langle \phi_1, \phi_2 \rangle^{}_\mathrm{ren} &= \langle \phi_1, \phi_2 \rangle^{}_{{\cal M}} + (\log y + \kappa) \langle \phi_1, \phi_2 \rangle^{}_{\partial {\cal M}} &{\rm for }& \ \nu = 1~,
\end{align}
\end{subequations}
where the subscripts ${\cal M}$, ${\partial {\cal M}}$ indicate the slices in which the usual KG products are to be evaluated.
As outlined in section~\ref{sec:KGfield}, after inserting the mode decomposition $\phi = Y f$ in (\ref{ren ip}) we find
\begin{equation}\label{ren ip modes}
    \langle \phi_1, \phi_2 \rangle^{}_\mathrm{ren} = \langle Y_{\sigma_1, \vec{j}_1}, Y_{\sigma_2, \vec{j}_2} \rangle^{}_\mathrm{slice} \langle f_1, f_2 \rangle^{}_\mathrm{SL, ren}~,
\end{equation}
where
\begin{subequations}
\begin{align}\label{SL ren}
    \langle f_1, f_2 \rangle_\mathrm{SL, ren}  &= \langle f_1, f_2 \rangle_\mathrm{SL} - \frac{(\sinh R)^{d-2}}{2(\nu-1)} f_1^* f_2 \big|_{R = \infty}  & {\rm for } \ \nu \neq 1~,
\\ \label{SL ren nu=1}
    \langle f_1, f_2 \rangle_\mathrm{SL, ren}  &= \langle f_1, f_2 \rangle_\mathrm{SL} + ( \kappa + \log 2 - R ) \frac{(\sinh R)^{d-2}}{2(\nu-1)} f_1^* f_2 \big|_{R = \infty}  & {\rm for } \ \nu = 1~.
\end{align}
\end{subequations}
Note that the first factor in the right hand side of (\ref{ren ip modes}) corresponds to the inner product of two dS harmonics with
different values of $\sigma$. At first sight, this might seem problematic since the inner product (\ref{dS ip}) was only defined for
two harmonics with the same boundary mass. However, as we shall see shortly, the renormalized SL factor defined in (\ref{SL ren})
vanishes for $\sigma_1 \neq \sigma_2$, so no inconsistency arises. Finally, as mentioned in section~\ref{sec:KGfield},
we note that the unrenormalized SL product can be evaluated by means of (\ref{eqn:reducedSL}) with $\lambda = (\sinh R)^{d-2}$.

We shall explicitly verify below that the inner product (\ref{ren ip modes}) is finite and conserved in the cases of interest, namely,
Dirichlet boundary conditions for all $\nu$ and Neumann boundary conditions for non-integer $\nu$ in the range $0 < \nu < 2$.

\subsection{Beyond the unitarity bound}

Let us now study the ghost content of the theories defined by the boundary conditions of interest. In order to do so,
we shall first find the spectrum of normalizable solutions and then compute the norms of the various modes.
We focus on the requirement of normalizability in the interior, i.e. $R=0$ in the coordinate system (\ref{eqn:ads-ds}),
since normalizability at the boundary is either automatic (for Dirichlet boundary conditions and Neumann boundary conditions
for $0< \nu < 1$) or guaranteed by the presence of the boundary terms (for Neumann boundary conditions and $\nu \geq 1$).
In the present section we restrict ourselves to non-integer $\nu$ and postpone the analysis of the special case $\nu = 1$ until
section~\ref{nu=1 dS}.

As stated in section~\ref{sec:KGfield}, the equation of motion is given by (\ref{eqn:KG}) with $\lambda = \sinh R$.
Using the mode decomposition and the property (\ref{dS harmonics}), this reduces to (\ref{eqn:radial-eq})
with $M^2 = - \sigma(\sigma + d - 1)$, which, according to the general discussion in section~\ref{sec:KGfield},
can be written as a SL problem with eigenvalue $\alpha = - \sigma(\sigma + d - 1)$.
Studying the radial equation near $R = 0$, we find that the two characteristic behaviors are $f \approx R^\sigma$
and $f \approx R^{1-d-\sigma}$. Inspecting (\ref{ren ip}), we conclude that for $\Re  \sigma > - (d-1)/2$ only $f \approx R^\sigma$ is
normalizable near the origin, while for $\sigma$ in the principal series, both fall-offs are $\delta$-function normalizable at the
horizon\footnote{By this we mean that they oscillate near $R=0$ in such a way that we can construct wave packets that decay faster than
any power law. Instead of constructing these wave packets, one can treat the norms of the modes in the principal series as distributions.
We shall do so below and obtain well-defined results. As anticipated above, this resembles the behavior of time-like modes
near the Poincar\'e horizon of Poincar\'e AdS.}.

In order to write down the full solution, we introduce $x = (\cosh R)^{-1}$, so the boundary is located at $x=0$ while the deep interior
corresponds to $x=1$. In terms of this variable, the two independent solutions can be expressed as
\begin{equation}\label{fDN dS}
    f_\mathrm{D/N} = x^{d/2+a^{}_\mathrm{D/N}\nu} (1 - x^2)^{\sigma/2} ~ _2F_1 \Big( c^{}_\mathrm{D/N}\,,\, c^{}_\mathrm{D/N} + \frac{1}{2} \,;\, 1 + a^{}_\mathrm{D/N}\nu \,,\, x^2  \Big),
\end{equation}
where $c^{}_\mathrm{D/N} = (d + 2 \sigma + 2 a^{}_\mathrm{D/N} \nu)/4$ and $a^{}_\mathrm{D} = 1$, $a^{}_\mathrm{N} = -1$.
Near the boundary, the radial profiles \eqref{fDN dS} behave as
\begin{equation}\label{bndy asympt}
    f_\mathrm{D} = x^{d/2+\nu}(1 + O(x^2))~, \qquad f_\mathrm{N} = x^{d/2-\nu}(1 + O(x^2))~,
\end{equation}
where the sub-leading terms consist solely of integer powers of $x$. Noting that near the boundary we have $x = y + O(y^3)$ and
comparing (\ref{bndy asympt}) with (\ref{eq:asymptSol}), we conclude that the profile $f_\mathrm{D}$ satisfies Dirichlet boundary conditions
while $f_\mathrm{N}$ satisfies Neumann boundary conditions.

It is convenient to organize the following discussion according to the value of $\sigma$ that characterizes the radial profiles.
For $\sigma$ in the principal series, we have seen that both characteristic behaviors are allowed near the origin.
This implies that the spectrum is continuous. In fact, Dirichlet/Neumann modes are simply given by the profiles \eqref{fDN dS}
with $\sigma = - (d-1)/2 + i \rho$. We now proceed to compute the respective norms, closely following \cite{Andrade:2011dg}.
We observe that since the spectrum is continuous, the norms must the thought of in the distributional sense.
Taking $\rho >0$ by convention, a simple calculation reveals that the symplectic products for the modes in the principal series are
\begin{equation}\label{DN ips}
    \langle \phi_1, \phi_2 \rangle^{}_\mathrm{ren} = \delta_{\vec{j}_1,\vec{j}_2} \delta(\rho_1 - \rho_2)  \left[\frac{\pi \Gamma(1+a_{D/N}\nu)}{2^\nu \rho \sinh(\pi \rho)}\right]^2 \bigg| \frac{1}{\Gamma(1- a_{D/N} \nu + i \rho) \Gamma(i \rho) } \bigg|^2~.
\end{equation}
It should be noted that the renormalized SL product yields the factor of $\delta(\rho_1 - \rho_2)$ in (\ref{DN ips}).
Thanks to this property, we only need to compute the dS inner product for modes of the same boundary mass.
Moreover, consistently with the fact that $\sigma$ belongs to the principal series, we have assumed the harmonics to be normalized as
$\langle Y_{\sigma, \vec{j}}, Y_{\sigma, \vec{k}} \rangle^{}_{{\rm slice}}  = \delta_{\vec{j},\vec{k}}$~.
Note that \eqref{DN ips} is positive definite for both Dirichlet and Neumann boundary conditions.
Also, we emphasize that, for $1< \nu < 2$ and Neumann boundary conditions, the explicit boundary contribution in (\ref{ren ip})
exactly cancels a divergence coming from the bulk term so that (\ref{DN ips}) is finite, as promised. Finally, we note
that \eqref{DN ips} does not mix modes of different quantum numbers, which guarantees conservation of the symplectic structure.

Let us consider now the case $\Re \sigma > - (d-1)/2$. With this restriction, only the solution that behaves
as $f \approx R^\sigma$ near $R = 0$ is normalizable, which implies that the allowed values of $\sigma$ form a discrete set.
In fact, expanding (\ref{fDN dS}) near $R = 0$, we find
\begin{equation}\label{fD orig}
    f_\mathrm{D/N} =2^{a^{}_\mathrm{D/N} \nu}\pi^{-\frac{1}{2}} \Gamma(1+a^{}_\mathrm{D/N}\nu)
     \Big(C^{(1)}_\mathrm{D/N} (R^\sigma+\dots) + C^{(2)}_\mathrm{D/N} (R^{1- d - \sigma}+\dots)\Big)~,
\end{equation}
where the ellipses denote subleading terms and
\begin{align}\label{C12 dS DN}
    C^{(1)}_\mathrm{D/N} &= \frac{2^{-\frac{1}{2}(d+\sigma)} \Gamma\left(\frac{1}{2}-\frac{d}{2}-\sigma \right)}{\Gamma\big(1-\frac{d}{2}+a^{}_\mathrm{D/N}\nu -\sigma \big)}~,
&
    C^{(2)}_\mathrm{D/N} &= \frac{2^{\frac{1}{2} (\sigma-1)} \Gamma\left(\frac{d-1}{2}+\sigma \right)}{\Gamma\big(\frac{d}{2}+a^{}_\mathrm{D/N}\nu +\sigma \big)}~.
\end{align}
As stated above, normalizability requires $C^{(2)}_\mathrm{D/N} = 0$, which translates into the quantization condition
\begin{equation}\label{sigma D}
    \sigma = \sigma^{}_\mathrm{D/N} := - \frac{(d-1)}{2} - n - \left(a^{}_\mathrm{D/N} \nu + \frac{1}{2} \right) \ \ \ \ {\rm for } \ n \in \mathbb{N} \cup \lbrace 0\rbrace~.
\end{equation}
Note that $\sigma_\mathrm{D}$ violates the assumption $\Re \sigma > - (d-1)/2$ for all $n$, so there are no Dirichlet modes in this class.
On the other hand, depending on the value of $\nu$, some discrete modes are allowed for Neumann boundary conditions.
In particular, restricting ourselves to $0 < \nu < 2$, we note that the mode $n = 0$ is allowed for $\nu > 1/2$, while $n= 1$
is allowed for $\nu > 3/2$.

Now, the norm of the Neumann modes that satisfy (\ref{sigma D}) is given by
\begin{equation}\label{norm principal N}
 \langle \phi_1, \phi_2 \rangle^{}_\mathrm{ren} =
   \langle Y_{\sigma_1, \vec{j}_1}, Y_{\sigma_2, \vec{j}_2} \rangle^{}_{{\rm slice}} \delta_{\sigma_1, \sigma_2} (-1)^{n}
   \frac{n! \csc(\pi \nu)}{2^{3+2n-2\nu}}\frac{(2\nu-2n-1)\Gamma\big(\nu-n-\frac{1}{2}\big)^2}{\Gamma(2\nu - n)}~,
\end{equation}
Note that for $1/2 < \nu < 1$ the $n  = 0$ mode has positive SL norm and $\sigma<0$, such that the overall norm is positive.
On the other hand, this mode has negative SL norm for $1< \nu < 2$, and also
the $n=1$ mode which belongs to the spectrum for $\nu > 3/2$ has negative SL norm.
Since the slice part $\langle Y_{\sigma_1, \vec{j}_1}, Y_{\sigma_2, \vec{j}_2} \rangle_{{\rm slice}}$
is positive for $\sigma<0$ and indefinite for $\sigma\geq 0$, we find ghosts in any case for $1<\nu<2$.

Summing up, we conclude that the Neumann spectrum is free of ghosts for $\nu < 1$, while for $\nu > 1$ the norm becomes indefinite.
In addition, the Dirichlet spectrum is ghost-free for all $\nu$, in complete agreement with the CFT unitarity bound.
It is worthwhile noting that we have found that it is possible to have a unitary theory that contains one
discrete --yet degenerate-- mode if the dimension of the operator is above but sufficiently close to the unitarity bound, i.e. for $d/2-1 < \Delta_- < d/2 - 1/2$.
Assuming that $\Delta_- = d/2 - \nu > 0$, so that the boundary operator is relevant, it follows from \eqref{sigma D} that this discrete mode always
occurs in the complementary series, which is in principle a continuous series. Interestingly enough,
the authors of \cite{Marolf:2010zp, Marolf:2010nz} encountered an analogous structure in the corrected 2-point function 
in the weakly interacting scalar theory in dS$_d$.
We can partly understand this qualitative agreement between the strongly and weakly coupled regimes
from the argument that operators close to the unitarity bound should interact weakly,
since operators saturating the unitarity bound must be free fields in a unitary theory.

\subsection{Saturating the unitarity bound}
\label{nu=1 dS}

So far we have assumed that $\nu$ is not an integer. For completeness, in this section we will tackle the case $\nu = 1$ and pay
special attention to the Neumann-like boundary condition (\ref{Neu nu=1}), although we shall also include the Dirichlet results.
The main feature of the integer $\nu$ cases is the presence of logarithms of the radial coordinate in the asymptotic expansion.
As a result, the counterterms required to renormalize the action contain the radial variable explicitly so conformal invariance
is broken, see (\ref{action dS n=1}).
The intuition about the existence of ghosts developed in the conformally invariant set-ups does therefore not transfer straightforwardly to this case.

We first proceed to find the spectrum of normalizable solutions. Again, we use the mode decomposition $\phi = Y f$, where $Y$ is a
dS harmonic and $f$ a radial profile. Introducing the variable $x =(\cosh R)^{-1}$, so that the boundary is at $x=0$, the two
independent solutions to the radial equation read
\begin{align}\label{ftilde1}
   {f}_1 &= x^{d/2-1}(1-x^2)^{\sigma/2} ~ _2F_1 \Big( \frac{\sigma-1}{2}+\frac{d}{4}\,,\, \frac{\sigma}{2}-\frac{d}{4}\,;\, \frac{d+1}{2}+\sigma\,,\, 1-x^2 \Big)~,
\\ \label{ftilde2}
   {f}_2 &= x^{d/2-1}(1-x^2)^{(1-d-\sigma)/2} ~ _2F_1 \Big( - \frac{d}{4}-\frac{\sigma}{2}\,,\, \frac{1-\sigma}{2}-\frac{d}{4}\,;\, \frac{3-d}{2}-\sigma\,,\, 1-x^2 \Big)~.
\end{align}
As in the non-integer $\nu$ case, both characteristic behaviors are allowed near the origin for modes in the principal series.
Thus, for both Dirichlet and Neumann boundary conditions, there are continuous families of  modes in the principal series and
one can readily verify that the norms are positive definite in this subspace.

Let us now study the discrete part of the spectrum. Defining $\sigma = -(d-1)/2 + \lambda$, the candidates for discrete modes are
those such that $\Re \lambda > 0 $. This is because in that case only (\ref{ftilde1}) is regular at the origin, which implies that
the boundary conditions at the conformal boundary will fix the value of $\lambda$. Letting $y = 2 e^{-R}$, we find that the
near-boundary expansion of (\ref{ftilde1}) is of the form (\ref{eq:asymptSol}) with
\begin{subequations}
\begin{align}\label{phiD2 n= 1}
    {f}_{1\mathrm{D}}^{(0)} &= \frac{2^{\lambda + \frac{1}{2}} \Gamma(1+\lambda)}{\sqrt{\pi}\, \Gamma\left(\lambda + \frac{3}{2} \right)}~,
         \qquad\qquad {f}_{1\mathrm{D}}^{(2)} = \frac{2^{\lambda -\frac{1}{2}} \Gamma(1+\lambda)}{\sqrt{\pi}\, \Gamma\left(\lambda -\frac{1}{2} \right)}~,
\\ \label{phiN0 n= 1}
    {f}_{1\mathrm{N}}^{(0)} &=  \frac{{f}_{1\mathrm{D}}^{(0)}}{16} \big[2d - 4 + \left(4 \lambda ^2-1\right)( 2\psi(\lambda +1/2) + 2 \gamma - 1 - \log 4 ) \big]~,
\end{align}
\end{subequations}
where $\psi$ is the digamma function and $\gamma$ is the Euler-Mascheroni constant. Dirichlet boundary conditions
require ${f}_{1\mathrm{D}}^{(0)} = 0$, which, according to \eqref{phiD2 n= 1} imply $\lambda = -(n + 3/2)$, where $n$ is a non-negative
integer. Since this violates our assumption $\Re \lambda >0$ for all $n$, we conclude that there are no Dirichlet modes in this class.
We now study the discrete Neumann modes. It follows from (\ref{Neu nu=1}) that these must
satisfy $2 {f}^{(0)}_{1\mathrm{N}} + (1 - 2 \kappa) {f}^{(2)}_{1\mathrm{D}} = 0$. We note that $\lambda = 1/2$ is a solution only
for $d=2$. Now, assuming $\lambda \neq 1/2$ and given (\ref{phiD2 n= 1}), (\ref{phiN0 n= 1}), the Neumann condition  translates into
\begin{equation}\label{bc dS nu = 1}
  b(\lambda) :=   \frac{d-2}{4\lambda^2-1}- \tilde{\kappa} + \psi \left(\frac{1}{2}+\lambda \right)  = 0~,
\end{equation}
where $\tilde{\kappa} = \kappa - \gamma + \log 2$.
Though we have not found the spectrum in closed form, it is still possible to extract the relevant physics.
In order to do so,
we first recall that complex solutions constitute a pair of ghost/antighosts, so we only need to examine the norms of the real $\lambda$ solutions. Assuming that such solutions exist, the norm of the corresponding modes can be written as
\begin{equation}\label{norm disc nu=1}
    \langle \phi_1, \phi_2 \rangle^{}_{{\rm ren}} = \delta_{\vec{j}_1,\vec{j}_2} \delta_{\sigma_1, \sigma_2}\langle Y_{\sigma_1, \vec{j}_1}, Y_{\sigma_2, \vec{j}_2}  \rangle_{{\rm slice}} \langle f , f \rangle^{}_\mathrm{SL,ren}~,
\end{equation}
where
\begin{equation}\label{norm dS nu=1}
    \langle f, f \rangle^{}_\mathrm{SL,ren} = A(\lambda) (1-4 \lambda^2) \frac{d}{d \lambda} b(\lambda)
\end{equation}
with $ A(\lambda) = 4^{\lambda -1} \Gamma(\lambda) \Gamma(1+\lambda)/[\pi \Gamma\left(\frac{3}{2}+\lambda \right)^2] > 0$. In \eqref{norm dS nu=1}, $\lambda$ is given implicitly by the real solutions of (\ref{bc dS nu = 1}) that satisfy $\lambda > 0$. Note that in writing (\ref{norm disc nu=1}) we have not assumed that the dS harmonics belong to a unitary representation.

Let us now study the existence of solutions of \eqref{bc dS nu = 1}. To do so, we first note
that $b(0) = 2(1 - \log 2) - d - \gamma - \tilde{\kappa}$. In addition, we shall use
that $b \rightarrow - \infty$ as $\lambda \rightarrow 1/2^-$, along with the facts that $b \rightarrow + \infty$
when $\lambda \rightarrow 1/2^+$ and $b \rightarrow + \infty$ when $\lambda \rightarrow \infty$.
Given this, we note that if $\kappa < \kappa_{c,1} := 2 - d - 3 \log 2$, there is only one real solution, $\lambda_0$, that occurs
in the range $(0, 1/2)$. Moreover, this solution is such that $b'(\lambda_0) < 0$, so it follows from \eqref{norm dS nu=1} that
the associated mode is a ghost. If we increase $\kappa$ above $\kappa_{c,1}$, we find two real solutions in $(0,1/2)$ as long
as $\kappa < \kappa_{c,2}$, where $\kappa_{c_1} < \kappa_{c,2} < 0$. In this regime, the solution with higher value of $\lambda$ is a
ghost. Further increasing $\lambda$, the solutions move to the complex plane.
Finally, there is another threshold $\kappa_{c,3} > 0$
such that for $\kappa>\kappa_{c,3}$ there are real solutions in $(1/2, \infty)$.
To see this we note that $b(\lambda)\rightarrow \infty$ for $\lambda \rightarrow 1/2^+$ and for $\lambda \rightarrow \infty$.
It therefore has a minimum in $(1/2,\infty)$ with a minimum value $b_\mathrm{min}=b_\mathrm{min}\vert_{\tilde\kappa=0}-\tilde\kappa$.
For sufficiently large $\tilde\kappa$ the minimum value is negative and we thus find real solutions.
For at least one of them we have $\frac{d}{d\lambda}b>0$, such that it has negative SL norm.
Since $\langle Y_{\sigma, \vec{j}}, Y_{\sigma, \vec{j}}  \rangle_{{\rm slice}}$ is either positive for $\sigma<0$
or indefinite for $\sigma\geq 0$ this means we have ghosts in any case.

In summary, we have established analytically that theories with Dirichlet boundary conditions have a ghost-free spectrum
for $\nu = 1$. On the other hand, for the family of Neumann-like boundary conditions we have considered,
we have found that there are ghosts
for all values of $\kappa$.

\section{Conclusions}
\label{sec:conclusions}

In this article we have studied unitarity violations in CFTs defined on the maximally symmetric
dS and AdS spacetimes from a holographic perspective.
For this purpose we have considered a scalar field
on AdS$_{d+1}$ conformally compactified such that the conformal boundary is (A)dS$_d$.
The mass and boundary conditions on the AdS$_{d+1}$ conformal boundary were chosen
such that the bulk theories provide a dual description of a CFT that contains an operator violating known unitarity bounds,
i.e.\ $m^2\geq m_\mathrm{BF}^2+1$ and Neumann$_{d+1}$.

Starting with the case of AdS$_d$ on the boundary, we have adapted the well-known procedure
of holographic renormalization to this setting and found that the qualitative features of
the bulk theory strongly depend on the choice of boundary conditions on the AdS$_d$ boundary.
While the Dirichlet$_d$ boundary condition yields a full set of normalizable modes, choosing
Neumann$_d$ drastically reduces the spectrum.
For Dirichlet$_d$ we have found that for even and odd $\mathbb Z_2$ parity the spectrum of the
bulk theory contains ghosts for $m_\mathrm{BF}^2+1<m^2<m_\mathrm{BF}^2+2$ combined with Neumann$_{d+1}$.
For $m^2=m_\mathrm{BF}^2+1$ with Neumann$_{d+1}$ boundary condition,
which corresponds to a CFT with an operator saturating the unitarity bound, we
have also shown that ghosts are present in the bulk theory, although without obtaining
the spectrum in closed form.
Thus, we have found  that the non-unitarity of the dual CFT is well reflected in the bulk theory for
the Dirichlet$_d$ cases.
As argued in the main text, it is also possible to extrapolate our results to higher values of the bulk
mass even in the absence of an explicit expression for the renormalized action and inner products.
This has shown that also for higher values of $\nu$ the boundary non-unitarity is recovered in the bulk theory.
For Neumann$_d$ on the other hand, we have also found ghosts in the spectrum for Neumann$_{d+1}$
and $m_\mathrm{BF}^2+1<m^2<m_\mathrm{BF}^2+2$, but extrapolating our results to higher values of the
bulk mass we have found that in
certain cases Neumann$_d$ boundary conditions yield -- contrary to
expectations based on the unitarity bound -- a ghost-free spectrum.
This can be traced back to the special structure of the boundary theory,
which lacks conformal invariance.
Summing up, we find that the boundary unitarity bound is well reflected in the bulk theories in the
cases where it is expected to hold.

It is interesting to compare these results in more detail with the expectations based on the field
theory reasoning.
On the one hand,
the presence of the boundary breaks the symmetry group from SO$(2,d)$ to SO$(2,d{-}1)$, such that
one might expect the relevant unitarity bound to be that of $d{-}1$ dimensions.
On the other hand, for observables localized away from the boundary the relevant unitarity bound should
still be the $d$-dimensional one.
Thus, for degrees of freedom that are not confined to the boundary,
we still expect the $d$-dimensional unitarity bound to be relevant\footnote{We thank
David Berenstein for clarifying this point to us.}.
Our results, which state that the relevant bound is the $d$-dimensional one,
are in good agreement with this picture, as
we have not included degrees of freedom that solely reside on the boundary of
AdS$_d$, which could however be done along the lines of \cite{Compere:2008us}, \cite{Takayanagi:2011zk}.

The result that choosing Neumann$_d$ boundary conditions effectively
reduces the bulk theory to a boundary theory in a trivial way allows for some insight on the
possibility of multi-layered holographic dualities. More concretely,
suppose we start with a (super-)gravity theory on AdS$_{d+1}$ with Neumann$_{d+1}$ boundary conditions,
such that the boundary theory is a gravitational theory on AdS$_d$ \cite{Compere:2008us}.
This AdS$_d$ theory should again have a dual description in
terms of a $(d{-}1)$-dimensional theory on the  boundary of AdS$_d$.
The AdS$_{d+1}$ theory we started with would then be dual to
the $(d{-}1)$-dimensional theory `on the boundary of the boundary'.
In principle, one might imagine iterating this procedure even further,
by choosing on the AdS$_d$ slices coordinates such that their boundary is AdS$_{d-1}$,
and Neumann$_d$ boundary conditions.
This would be expected to relate the $(d{+}1)$-dimensional theory to a $({d{-}2})$-dimensional one.
However, for our construction the arguments for the dimensional reduction of the AdS$_{d+1}$ theory 
with Neumann$_d$ boundary condition
to a $d$-dimensional theory apply, making the further iterations trivial.
We conclude that a $(d{+}1)$-dimensional theory may be related to a $(d{-}1)$-dimensional one,
but that non-trivial relations between theories with spacetime dimensions differing by more than
two can not be obtained in that way.

For the case of dS$_d$ on the boundary the involved geometry is an open patch of global AdS$_{d+1}$,
bounded by a causal horizon.
Although the setup is similar to Poincar\'{e} AdS in that respect, we found -- in contrast to
Poincar\'{e} AdS -- a straightforward reflection of the boundary non-unitarity since the spectrum of the bulk theory contains
ghosts.
The difference in the two settings is that in our setup the dS$_d$ slices have compact spatial
sections, which is different from Poincar\'{e} AdS where the $d$-dimensional slices are Minkowski.
This suggests that the tricky manifestation of the boundary non-unitarity in the bulk found for
Poincar\'{e} AdS is related to the non-compactness of the  boundary, rather than to the appearance
of a horizon in the bulk.
To further investigate this point one could study the case with dS on the boundary using
an open slicing instead of global dS$_d$ coordinates.

We have also included the cases with Dirichlet boundary conditions on the conformal boundary of AdS$_{d+1}$
for generic\footnote{Note that the results for Dirichlet$_{d+1}$ are insensitive to
the $\nu$-dependent explicit form of the counterterms due to the fast fall-off of the field.} $\nu$,
and the fact that we found ghost-free spectra in that case shows that the condition $\Delta\geq d/2-1$, derived
in \cite{Minwalla:1997ka} as necessary condition for unitarity, is indeed also sufficient for CFTs which have a holographic description
in terms of the setups we have considered.
In summary, our results
show that the (non-)unitarity of the boundary CFTs is well reflected in the dual bulk theories.

%%%%%%%%%%%%%%%%%%%%%%%%%%%%%%%%%%%%%%%%%%%%%%%%%%%%%%%%%%%%%%%%%%%%%%%%
\section*{Acknowledgements}

We thank David Berenstein, Ian Morrison, Thorsten Ohl and Alexander Schenkel for useful conversations and correspondence.
We are specially grateful to Don Marolf for providing numerous insights and for reviewing an earlier version of this manuscript.
CFU is pleased to thank University of California, Santa Barbara and in particular the gravity group
for their kind hospitality during the initial stages of this work.
TA is pleased to thank University of California, Davis for their hospitality during the completion of this work.
TA was partly supported by a Fulbright-CONICYT fellowship, by the US National Science Foundation under grant PHY08-55415 and
by funds from the University of California.
CFU is supported by the German National Academic Foundation
(Studienstiftung des deutschen Volkes) and by Deutsche Forschungsgemeinschaft through the Research Training Group GRK\,1147
\textit{Theoretical Astrophysics and Particle Physics}.

\appendix

\section{Below the BF bound in global AdS} \label{app:tachyon-globalAdS}

In section~\ref{sec:saturateUnitarityAdS} we found tachyonic transverse modes below the BF bound,
which we discuss in more detail now.
We consider a scalar field $\phi$ with squared mass $m^2 = - d^2/4 + (i\lambda)^2$, $\lambda\in\mathbb R$
below the BF bound
on global AdS$_{d+1}$ with metric
\begin{equation}
    ds^2 = \sec^2 \rho (- dt^2 + d \rho^2) +  \tan^2 \rho d \Omega_{d-1}~.
\end{equation}
Changing the radial coordinate to $r= \cos \rho$ such that the boundary is located at $r  = 0$,
the asymptotic expansion of the field reads
\begin{equation}\label{eqn:asympt}
    \phi = r^{d/2 + i \lambda} \phi^\sbr{+} + r^{d/2 - i \lambda} \phi^\sbr{-}~.
\end{equation}
The symplectic structure constructed from the symplectic current
$\omega^\mu(\phi_1, \phi_2) = i g^{\mu \nu}( \phi_1 \partial_\nu \phi_2 - \phi_2 \partial_\nu \phi_1 )$
is finite without adding counterterms, and we shall impose
boundary conditions at the conformal boundary such that it is conserved.
The flux through the boundary is given by
\begin{equation}\label{eqn:flux}
    \mathcal F = 2 i \lambda \int_{\partial M}\Big( \phi^\sbr{-}_1 \phi^\sbr{+}_2 - \phi^\sbr{-}_2 \phi^\sbr{+}_1 \Big)~.
\end{equation}
We choose a boundary condition which makes $\mathcal F$ vanish and
is compatible with reality of $\phi$ as a formal power
series\footnote{The boundary condition \eqref{eqn:bc} can be generalized to include a phase as
$\phi^{(+)}\big\vert_{r=0}\:=\:  e^{i 2 \alpha} \phi^{(-)}\big\vert_{r=0}$, $\alpha \in \mathbb{R}$.
This corresponds to rescaling the coordinate $r$ as can be seen from \eqref{eqn:asympt},
and we therefore set $\alpha=0$ without loss of generality.}
\begin{equation}\label{eqn:bc}
    \phi^{(+)}\big\vert_{r=0}\:=\:  \phi^{(-)}\big\vert_{r=0}  ~.
\end{equation}
It should be noted that the boundary condition \eqref{eqn:bc} breaks
invariance under radial isometries, as it relates the coefficients of different
powers of $r$.
Alternatively, from the boundary perspective
conformal invariance is broken since the operators associated to $\phi^{(+)}$
and  $\phi^{(-)}$ have different conformal dimensions.

In order to solve the Klein-Gordon equation, we use the mode decomposition
$\phi = e^{- i \omega t} Y_{L}(\Omega) \psi(r)$
where $Y_L$ is a spherical harmonic on $S^{d-1}$ satisfying $\Delta^{}_{\Omega_{d-1}}Y_L=-L(L+d-2) Y_L$.
For notational convenience we introduce
$a_\pm:=c\pm \frac{\omega}{2}$ and $b_\pm:=c^\ast\pm \frac{\omega}{2}$, where
$c=(d+2L-2i\lambda)/4$.
For $\lambda\in\mathbb R$ the solution which is regular at the origin is (see e.g.\ \cite{Andrade:2011dg})
\begin{equation}\label{eqn:psi1}
  \psi(r) = r^{\frac{d}{2}- i\lambda } \big(1-r^2\big)^\frac{L}{2}\,
   _2F_1 \Big(a^{}_-\,,\,a^{}_+\,,\,\frac{d}{2}+L\,,\,1-r^2\Big)~.
\end{equation}
Note that, using
$_2F_1(a,b;c;z) = (1-z)^{c-a-b} ~ _2F_1(c-a,c-b;c;z)$,
one can show that the radial profile (\ref{eqn:psi1}) is real for $\omega^\ast = \pm\omega$.
From (\ref{eqn:psi1}) we find $\phi^\sbr{\pm}\big\vert_{r=0}=e^{- i \omega t} Y_{L}(\Omega) \psi^{}_\pm$ where
\begin{equation}\label{eqn:phipm}
    \psi^{}_+ = \frac{\pi {\rm csch }(\pi  \lambda)\Gamma\left(\frac{d}{2}+L\right)}{\lambda\,\Gamma(i \lambda)
     \Gamma\left(a_-\right) \Gamma\left(a_+\right)}~,\qquad
    \psi^{}_- = \frac{\pi  {\rm csch}(\pi  \lambda) \Gamma\left(\frac{d}{2}+L\right)}{\lambda\,\Gamma(-i \lambda)
     \Gamma\left(b_-\right) \Gamma\left(b_+\right)}~.
\end{equation}
The boundary condition (\ref{eqn:bc}) therefore amounts to $\psi^{}_+=\psi^{}_-$. This is equivalent to\footnote{
The $\Gamma$-functions in the denominator only have poles or zeros if $\Im(\omega)=\pm\lambda$.
This, however, does not yield solutions since due to the structure of the arguments
the pole/zero always appears in one of $\psi^{}_\pm$ only.
}
\begin{equation}\label{eqn:bc2}
 \frac{\Gamma(i \lambda)}{\Gamma(-i \lambda)}
=\frac{ \Gamma\left(b_-\right) \Gamma\left(b_+\right)  }{\Gamma(a_-) \Gamma(a_+)}     ~.
\end{equation}
We first show that there are only real or purely imaginary solutions.
Using the Weierstra{\ss} form $\Gamma(z)^{-1}=z e^{\gamma z}\prod_{k=1}^\infty (1+z/k)e^{-z/k}$,
the modulus of (\ref{eqn:bc2}) yields
\begin{equation}
 1=\left|\frac{\Gamma(b_+)\Gamma(b_-)}{\Gamma(a_+)\Gamma(a_-)}\right|^2
=\prod_{k=0}^\infty \left| \frac{(k+a_+)(k+a_-)}{(k+b_+)(k+b_-)}\right|^2
=\prod_{k=0}^\infty \left(1+\frac{(d/2+L+2k)\lambda\, \Re(\omega)\Im(\omega)}{\left|k+b_+\right|^2\,\left|k+b_-\right|^2}\right)~.
\end{equation}
The first equality follows from (\ref{eqn:bc2}),
the second one by using the Weierstra{\ss} form
and the third one by evaluating each factor.
Depending on the sign of $\lambda\, \Re(\omega)\Im(\omega)$,
either each factor in the product is greater than one, or each factor is less than one.
As that makes the entire product on the right different from $1$,
we conclude that there are no solutions with $\Re(\omega)\neq 0$ and $\Im(\omega)\neq 0$.

For $\omega$ real or purely imaginary the modulus of both sides of (\ref{eqn:bc2}) is identically $1$.
We first analyze purely imaginary $\omega$.
In this case we can use the asymptotic expansion
$\Gamma(z)=\sqrt{2\pi}e^{-z}z^{z-1/2} \big(1+\mathcal O(|z|^{-2})\big)$
which holds if $z$ is bounded away from the negative real axis ($\exists\, \delta>0:\ |\arg z|<\pi-\delta$).
Parametrizing $\omega=i(\lambda+2e^\tau)$ we find
\begin{equation}\label{eqn:bc2appr1}
 \frac{ \Gamma\left(b_-\right) \Gamma\left(b_+\right)  }{\Gamma(a_-) \Gamma(a_+)}
  = e^{-2i\lambda}\sqrt{\frac{a_+a_-}{b_+b_-}}\,\frac{b_-^{b_-}b_+^{b_+}}{a_-^{a_-}a_+^{a_+}}+\mathcal O(|\omega|^{-2})
 = e^{2i\lambda\tau}+\mathcal O(e^{-2\tau})~.
\end{equation}
The second equality follows from the asymptotic expansion for generic large $\omega$
and the third one by expanding the result for large imaginary part, i.e.\ large $e^\tau$.
Therefore, in the regime of large $\tau$, solving (\ref{eqn:bc2}) becomes equivalent to solving
$2\lambda\tau=2\arg\Gamma(i\lambda) \mod 2\pi$.
This yields a discrete series of solutions which for large $|\omega|$ is well approximated by
$\omega=\pm i(\lambda+2e^\tau)$ with $\tau=\lambda^{-1}(\arg\Gamma(i\lambda)+\pi k)$, $k\in\mathbb Z$.
Note that for $\lambda \rightarrow 0$ the imaginary frequency solutions go off to $\pm i \infty$,
consistent with the fact that there are no complex solutions for $\lambda = 0$.
We stress that the presence of these imaginary frequency solutions indicates the expected instabilities that are known to occur for
masses below the BF bound. Moreover, as argued in \cite{Andrade:2011dg}, the imaginary frequency solutions constitute a
pair `ghost/anti-ghost'.

For the real solutions we assume without loss of generality $\omega>0$.
Using $\Gamma(z)\Gamma(1-z)=\pi/\sin(\pi z)$ to rewrite (\ref{eqn:bc2})
such that all the arguments of the $\Gamma$ function have positive real part
and then using the asymptotic expansion discussed above yields
\begin{equation}\label{eqn:bc2appr2}
\frac{ \Gamma\left(b_-\right) \Gamma\left(b_+\right)  }{\Gamma(a_-) \Gamma(a_+)}
=\frac{\sin(\pi a_-)}{\sin(\pi b_-)}\frac{\Gamma(b_+)\Gamma(1-a_-)}{\Gamma(a_+)\Gamma(1-b_-)}
=\Big(\frac{\omega}{2}\Big)^{2 i\lambda}\,\frac{\sin\big(\frac{\pi}{2}(\omega+i\lambda)\big)}{\sin\big(\frac{\pi}{2}(\omega-i\lambda)\big)}+\mathcal O(\omega^{-1})~.
\end{equation}
Equation (\ref{eqn:bc2}) for large $\omega$ then becomes
\begin{equation}\label{eqn:bc2real}
\frac{\Gamma(i \lambda)}{\Gamma(-i \lambda)}
=\Big(\frac{\omega}{2}\Big)^{2 i\lambda}\,\frac{\sin\big(\frac{\pi}{2}(\omega+i\lambda)\big)}{\sin\big(\frac{\pi}{2}(\omega-i\lambda)\big)}
=e^{ 2i\big(\lambda\log\frac{\omega}{2}+\arctan (\tanh\frac{\pi\lambda}{2}\cot\frac{\pi\omega}{2})\big)}=:e^{i \vartheta(\omega)}~.
\end{equation}
While $e^{i \vartheta(\omega)}$ is of course single-valued,
the $\arctan$ is single-valued only up to addition of integer multiples of $\pi$.
We choose the values within these classes such that $\arctan (a\cot\frac{\pi\omega}{2} )$
becomes a continuous function on $\mathbb R$,
e.g.\ $\arctan (a\cot\frac{\pi\omega}{2} ) =\arctan_0 (a\cot\frac{\pi\omega}{2} ) - \mathrm{sign}(a)\pi \big\lfloor\frac{\omega}{2}\big\rfloor$,
where $\lfloor x\rfloor$ denotes the greatest integer smaller than $x$
and $\arctan_0$ is the principal value in $[-\frac{\pi}{2},\frac{\pi}{2}]$.
This makes $\vartheta$ a continuous function which tends to $-\infty$ for $\omega\rightarrow \infty$.
Due to the periodicity of $e^{i\vartheta(\omega)}$ this shows that there is a series of solutions to (\ref{eqn:bc2real}).
Thus, we also have a series of real solutions to (\ref{eqn:bc2}).

Finally, we note that the approximations derived above in (\ref{eqn:bc2appr1}), (\ref{eqn:bc2appr2})  provide
an accurate description already for moderately large arguments of the $\Gamma$-functions.

\bibliographystyle{JHEP-2.bst}
\bibliography{adsads}

\end{document}